\documentclass[twocolumn]{aastex63}
\usepackage[caption=false]{subfig}
\usepackage{mathtools,amssymb,CJK, gensymb}
\newcommand{\vsh}{v_{\rm sh}}

\newcommand{\nism}{n_{\rm ISM}}

\newcommand{\w}[1]{v_{\rm A,#1}}

\newcommand{\Emax}{E_{\rm max}}

\newcommand{\gth}{\gamma_{\rm th}}
\newcommand{\gcr}{\gamma_{\rm CR}}

\submitjournal{ApJ}

\shorttitle{Nonthermal Signatures of Radiative Supernova Remnants II}

\begin{document}

\title{Nonthermal Signatures of Radiative Supernova Remnants II: \\ The Impact of Cosmic Rays and Magnetic Fields}

\begin{CJK*}{UTF8}{gbsn}

\correspondingauthor{Rebecca Diesing}
\email{rrdiesing@ias.edu}

\author[0000-0002-6679-0012]{Rebecca Diesing}
\affiliation{School of Natural Sciences, Institute for Advanced Study, Princeton, NJ 08540, USA}
\affiliation{Department of Physics and Columbia Astrophysics Laboratory, Columbia University, New York, NY 10027, USA}

\author[0000-0002-1030-8012]{Siddhartha Gupta}
\affiliation{Department of Astrophysical Sciences, Princeton University, Princeton, NJ 08540, USA}
\begin{abstract}

Near the ends of their lives, supernova remnants (SNRs) enter a ``radiative phase," when efficient cooling of the postshock gas slows expansion. Understanding SNR evolution at this stage is crucial for estimating feedback in galaxies, as SNRs are expected to release energy and momentum into the interstellar medium near the ends of their lives. A standard prediction of SNR evolutionary models is that the onset of the radiative stage precipitates the formation of a dense shell behind the forward shock. In Paper I, we showed that such shell formation yields detectable nonthermal radiation from radio to $\gamma$-rays, most notably emission brightening by nearly two orders of magnitude. However, there remains no observational evidence for such brightening, suggesting that this standard prediction needs to be investigated. In this paper, we perform magneto-hydrodynamic simulations of SNR evolution through the radiative stage, including cosmic rays (CRs) and magnetic fields to assess their dynamical roles. 
We find that both sources of nonthermal pressure disrupt shell formation, reducing shell densities by a factor of a few to more than an order of magnitude. We also use a self-consistent model of particle acceleration to estimate the nonthermal emission from these modified SNRs and demonstrate that, for reasonable CR acceleration efficiencies and magnetic field strengths, the nonthermal signatures of shell formation can all but disappear. We therefore conclude that the absence of observational signatures of shell formation represents strong evidence that nonthermal pressures from CRs and magnetic fields play a critical dynamical role in late-stage SNR evolution. \\

\end{abstract}

\section{Introduction} \label{sec:intro}

Supernova remnants (SNRs) play a critical role in galaxy formation and evolution by injecting energy and momentum into the interstellar medium (ISM). This injection can drive large-scale winds which quench star formation and enrich the intergalactic medium with metals \citep[e.g., ][]{ht17}. To model these effects, galaxy formation simulations rely on subgrid prescriptions for SNR ``feedback" \citep[e.g., ][see also \cite{crain+23} for a recent review]{agertz+13,hopkins+18, pfrommer+17,pillepich+18,feldmann+23}. 

In order to properly tune these feedback prescriptions, SNR evolution has been studied in a variety of environments \citep[e.g.,][]{chevalier+74, kim+15}. While the precise nature of SNR evolution has been shown to depend on the properties of the ambient interstellar medium (ISM), the life of an SNR can be broadly separated into three main phases: I) a free-expansion, or ``ejecta-dominated" phase \citep[e.g.,][]{chevalier82, truelove+99}, which persists until the swept-up mass becomes comparable to the ejecta mass; II) an adiabatic, or ``Sedov-Taylor" phase \citep[][]{sedov59, taylor50}, which continues until the shock slows such that the postshock temperature drops below $\sim 10^6$ K and cooling becomes efficient; and III) a ``radiative" phase, in which thermal gas loses its energy via atomic transitions and expansion continues first due to the SNR's internal pressure and then due to momentum conservation \citep[e.g., ][]{cioffi+88, draine11, bandiera+10}.

One of the key predictions of this evolutionary picture is the formation, at the onset of the radiative phase, of a dense shell behind the shock, resulting from the drop in thermal pressure support. \citep[e.g.,][]{ostriker+88,bisnovatyi-kogan+95}.
This shell formation occurs after roughly $1-5\times 10^4$ yr \citep[for typical ISM densities, e.g.,][]{kim+15} and, in the standard hydrodynamical picture, leads to density enhancement factors approaching $\sim 10^2$ relative to the ambient medium, depending on the shock Mach number \citep[see, e.g.,][for some useful estimates]{kim+15, diesing+24}. However, a definitive detection of such a shell is still missing. Searches have been conducted using neutral hydrogen emission, but only partial shells have been reported in the literature \citep[see, e.g.,][]{koo+20}.

Radiative shells may also be detectable via their nonthermal emission. Namely, SNRs are expected to efficiently accelerate nonthermal particles, or cosmic rays \citep[CRs, e.g.,][]{hillas05,berezhko+07,ptuskin+10,caprioli+10a} at their forward shocks via diffusive shock acceleration \citep[DSA, e.g.,][]{fermi54, krymskii77, axford+77p, bell78a, blandford+78}. These CRs then interact with postshock material, producing emission that extends from radio to $\gamma$-rays. This emission has been observed extensively in young SNRs \citep[e.g.,][]{morlino+12, slane+14, ackermann+13}, which are expected to accelerate the largest number of particles and, in the absence of shell formation (or dense ambient media), be brighter than their older counterparts. However, if the standard picture of SNR evolution holds, the presence of a dense shell can produce a dramatic rise in nonthermal emission at late evolutionary stages (\cite{diesing+24}, see also \cite{lee+15, brose+20, kobashi+22}). More specifically, the compressed magnetic fields and enhanced densities inside the shell provide excellent targets for synchrotron and proton-proton interactions. 

In Paper I \citep{diesing+24}, we investigated this nonthermal brightening and demonstrated that, in the standard picture of SNR evolution, the onset of the radiative phase is accompanied by radio and $\gamma$-ray enhancements approaching two orders of magnitude. These enhancements include appreciable TeV emission above a typical SNR's high-energy cutoff, meaning that the onset of the radiative phase should produce a $\gamma$-ray signal that will be readily detectable by the Cherenkov Telescope Array (CTA). However, there exists little to no observational evidence for this effect. While present-day $\gamma$-ray telescopes do not have sufficient resolution to distinguish between shells and molecular cloud interactions, the lack of a TeV detection of an SNR older than $t \sim 4 \times 10^3$ yr \citep{HESS18b} sheds doubt on a picture in which $\gamma$-ray emission is so enhanced that TeV emission can be detected above a high-energy cutoff. Even more concerningly, a complete shell--which is predicted to form even in the presence of a highly nonuniform medium \citep[][]{guo+24}--has yet to be detected in the radio. While radiative SNRs such as W44 \citep[][]{castelletti+07} and IC443 \citep[][]{castelletti+11} exhibit partial radio shells, the fact that they are incomplete indicates that these ``shells" may simply be the result of molecular cloud interactions. Meanwhile, a recent survey of 36 Galactic SNRs observed with the MeerKAT array revealed at most tenuous evidence for complete shell formation; the vast majority of the SNRs in the survey exhibit only partial shells, even in cases with relatively spherical geometry \citep[][]{cotton+24}. 

In short, observations point toward a picture in which shell formation is disrupted. As hydrodynamic instabilities appear insufficient to destroy radiative shells on the $\sim 10^5$ yr timescales considered in Paper I \citep{guo+24}, we instead look to shell disruption by nonthermal pressure contributions: CRs and magnetic fields. CRs, which can represent as much as $\sim 10-20$\% of the SNR bulk kinetic energy \citep[][]{caprioli+14a, park+15, gupta+24b}, are known to impact SNR evolution during the radiative stage \citep[][]{diesing+18}. Namely, as their energy is not radiated away at late times, they represent a reservoir of pressure available to continue SNR expansion and/or to prevent the formation of the dense shell. By extending the lives of their accelerators, CRs may meaningfully enhance SNR feedback. Similarly, magnetic fields are known to disrupt shell formation when oriented perpendicular to the shock normal; these perpendicular components are compressed, increasing the magnetic pressure in the vicinity of the shell \citep{petruk+18,sharma+14}. However, the combined effect of CR and magnetic pressure--particularly with realistic $\sim 3 \mu G$ ambient magnetic fields and a theoretically-motivated picture of CR transport--has not been investigated in the literature.

In this paper, we use two-fluid magneto-hydrodynamic (MHD) simulations to model the evolution of an SNR through the radiative phase, including the dynamical effects of both CRs and magnetic fields. We then quantify the impacts of CR acceleration and magnetic compression on shell formation, and couple these simulations with a semi-analytic model of particle acceleration to predict the nonthermal emission from SNRs with different CR acceleration efficiencies and magnetic field orientations.

This paper is organized as follows: in Section \ref{sec:method}, we describe our method, including our MHD simulation setup (\ref{subsec:hydro_method}) and our nonthermal emission calculations (\ref{subsec:emission_method}). We present our results in Section \ref{sec:results}, including our modeled SNR hydrodynamics (\ref{subsec:hydro_results}) and the observational consequences of CR and magnetic pressures (\ref{subsec:emission_results}). In Section \ref{sec:discussion}, we discuss the role of CRs and magnetic fields in modifying SNR feedback, particularly in the context of \cite{diesing+18}.
We summarize in Section \ref{sec:conclusion}.

\section{Method} \label{sec:method}

Herein we describe the model we use to explore the effects of CRs and magnetic fields on a typical SNR as it evolves from adiabatic to radiative. Throughout this work, we consider a representative case with initial energy $E_{\rm SN} = 10^{51}$ erg and ejecta mass $M_{\rm ej} = 1 M_{\odot}$, expanding into a uniform ISM with number density $\nism = 1$ cm$^{-3}$, temperature $T = 10^4$ K, solar metallicity, and perpendicular magnetic field (relative to the shock normal) $B_{\perp} = B_0 \sin{\theta} \in [0, 3] \mu$G. 

In short, our representative SNR is meant to approximate a Type Ia SNR expanding into the warm ionized phase of the ISM. However, as we focus exclusively on late-stage SNR evolution, we predict similar results for core-collapse SNRs, unless the progenitor star launched a strong stellar wind or formed in a cluster \citep[where stellar winds and previous SNe explosions drive different shock structures, e.g.,][]{yadav+17,gupta+18,gupta+20, el-badry+19}. Of course, the ISM itself is likely more complex than the medium considered in this work. We choose this simple, one-dimensional setup in order to isolate the dynamical impact of CRs and magnetic fields, with the understanding that real SNRs can and do exhibit more anisotropic evolution and emission.

\subsection{MHD simulations} \label{subsec:hydro_method}

To model shock evolution and shell formation, we conduct one-dimensional, spherically symmetric, ideal MHD simulations of a single supernova explosion without thermal conduction or mixing diffusion. As this work concerns radiative SNRs, we neglect the ejecta-dominated stage (and any associated explosion dynamics). More specifically, we  use the \texttt{PLUTO} code \citep{mignone+07}, modified to include CRs as an additional fluid component as described in detail in \cite{gupta+21a}, in order to solve the following coupled equations:

\begin{equation}
    \frac{\partial\rho}{\partial t} + \nabla \cdot (\rho\mathbf{v}) = 0,
\end{equation}
\begin{equation}
    \frac{\partial\rho\mathbf{v}}{\partial t} + \nabla \cdot (\rho\mathbf{v}\mathbf{v}-\mathbf{BB}+P_{\rm tot}) = 0,
\end{equation}
\begin{equation}
    \frac{\partial\mathbf{B}}{\partial t} + \nabla \cdot (\mathbf{v}\mathbf{B}-\mathbf{B}\mathbf{v}) = 0,
\end{equation}
\begin{equation}
    \frac{\partial E_{\rm tot}}{\partial t} + \nabla \cdot((E_{\rm tot}+P_{\rm tot})\mathbf{v}-(\mathbf{v \cdot B})\mathbf{B}) = -\rho L.
\end{equation}
\begin{equation}
    \frac{\partial E_{\rm CR}}{\partial t} + \nabla \cdot (E_{\rm CR}\mathbf{v}) = - P_{\rm CR} \nabla \cdot \textbf{v}.
\end{equation}
Here, $\rho$ is the mass density, $\mathbf{v}$ is the fluid velocity, $\mathbf{B}$ is the magnetic field, and $\rho L = n_{\rm H}^2\Lambda(T)$ is the cooling rate (where $n_{\rm H} \equiv \rho/(\mu m_{\rm p})$ is the hydrogen number density and $\mu = 0.6$ is the mean molecular weight of the ionized ISM). We use the same cooling curve as in Paper I and \cite{el-badry+19}, with a temperature floor set to $10^4$ K in order to approximate the balance between radiative cooling of the gas and heating driven by stellar or other radiation sources (e.g., \citealt{gupta+16}). $E_{\rm tot}$ and $P_{\rm tot}$ are the total energy density and pressure, given by,
\begin{equation}
    E_{\rm tot}  = \frac{1}{2}\rho|\mathbf{v}|^2 + \frac{P_{\rm th}}{\gamma_{\rm th} -1} + \frac{P_{\rm CR}}{\gamma_{\rm CR} -1} + \frac{|\mathbf{B}|^2}{2}
\end{equation}

\begin{equation}
    P_{\rm tot} = P_{\rm th }+ P_{\rm CR} + \frac{|\mathbf{B}|^2}{2},
\end{equation}
where $\gamma$ is the adiabatic index, equal to 5/3 for thermal gas and 4/3 for CRs. Throughout this work, subscripts ``th" and ``CR" denote thermal gas and CRs, respectively. 

Note that we neglect CR diffusion and consider only advection, which dominates transport for the $\sim$GeV particles contributing to the majority of the CR pressure. We also neglect CR cooling due to proton-proton losses, since the loss timescale even in dense environments ($ n_{\rm H} \simeq 10^2 \ \rm cm^{-3}$) far exceeds the SNR ages considered in this work. 

CRs are injected into shocked zones (identified using a method described in \citealt{pfrommer+17,gupta+21a}) to set the CR pressure fraction immediately downstream of the shock,
$\xi_{\rm CR} \equiv P_{\rm CR, 2}/(P_{\rm CR, 2} + P_{\rm th, 2})$. This prescription naturally decreases CR pressure injection during the radiative phase (i.e., when $P_{\rm th, 2}$ declines). Since shocks with low Mach numbers are expected to be inefficient particle accelerators \citep[e.g., ][]{caprioli+14a}, injecting CR pressure in this manner ensures that we do not overestimate the dynamical impact of CRs at late times (as we typically find Mach numbers $\gtrsim 10$, this estimate is actually quite conservative). Throughout this work, we consider $\xi_{\rm CR} = 0.0$, $\xi_{\rm CR} = 0.1$, and $\xi_{\rm CR} = 0.2$, which represent pessimistic, typical, and optimistic CR acceleration efficiencies based on the results of kinetic simulations \cite{caprioli+14a, caprioli+15}. In particular, the $\xi_{\rm CR} = 0.2$ case may approximate the case in which CR \emph{reacceleration} is efficient, as discussed in detail in, e.g., \cite{cardillo+16}, as well as in Paper I. Note that, in reality, acceleration efficiency depends strongly on the magnetic field orientation, $\theta$, with quasi-parallel ($\theta \lesssim 50^{\rm o}$) shocks achieving the highest $\xi_{\rm CR}$. We therefore focus our analysis on simulations with high (low) acceleration efficiencies and low (high) inclinations, emphasizing that scenarios in which $\xi_{\rm CR}$ and $\theta$ are both very large or very small are generally unphysical.

Throughout this work, we consider magnetic field orientations of $\theta = 0\degree$, $\theta = 45\degree$, and $\theta = 90\degree$ with respect to the shock normal. In principle, in our spherically symmetric, one-dimensional setup, any radial magnetic field would violate $\nabla \cdot \mathbf{B} = 0$. Conveniently, however, only the magnetic field components perpendicular to the shock normal (i.e., tangential to the shock surface) are dynamically important, since only these components are compressed. As such, we approximate these three orientations by setting only a perpendicular component of the magnetic field, $B_{\perp} = B_0\sin{\theta}$.

Since we are not concerned with the early evolution of the SNR, the SN explosion is initiated by depositing energy out to 0.1 pc. We use reflecting and outflow boundary conditions at the inner and outer boundaries, respectively. To ensure that we are properly resolving the structure behind the shock, we increase grid resolution until the maximum density of our radiative shell converges. Throughout this work, we show our highest-resolution runs, with fixed grid resolution $\Delta = 10^{-3}$ pc (note that this resolution is sufficient to resolve the radiative relaxation layer). 

\subsection{Nonthermal emission}
\label{subsec:emission_method}

In order to assess the observational consequences of nonthermal pressure during late-stage SNR evolution, we couple our MHD simulations with a detailed calculation of particle acceleration and nonthermal emission described below and in greater detail in Paper I. Namely, we model CR acceleration using a semi-analytic prescription that self-consistently solves the steady-state transport equation for the distribution of nonthermal particles at the forward shock, including the effects of magnetic field amplification and shock modification due to the presence of CRs. 
This model for estimating nonthermal emission is discussed in Paper I, see also 
\cite{caprioli+09a,caprioli+10b, caprioli12, diesing+19, diesing+21} and references therein, in particular \cite{malkov97,malkov+00,blasi02,blasi04,amato+05, amato+06}. 

We inject particles with momenta above $p_{\rm inj} \equiv \zeta_{\rm inj}p_{\rm th}$ into the acceleration process, where $p_{\rm th}$ is the thermal momentum and we choose $3.6 \leq \zeta_{\rm inj} \leq 3.7$ to produce CR pressure fractions comparable to those assumed in our MHD simulations. The exception is the $\xi_{\rm CR} = 0.0$ case which, if fully self-consistent, would produce no CRs and therefore no nonthermal emission. Instead, when $\xi_{\rm CR} = 0.0$, we set $\zeta_{\rm inj} = 3.7$ as in the $\xi_{\rm CR} = 0.1$ case to approximate the scenario in which CRs are being accelerated but have no dynamical impact on the SNR.

As in Paper I, we calculate the maximum proton energy by requiring that the diffusion length (assuming Bohm diffusion) of particles with energy  $E = \Emax$ be 5$\%$ of the shock radius. Furthermore, to ensure a conservative estimate of our SNR's TeV emission, we assume that diffusing (not escaping) CRs are responsible for magnetic field amplification \citep[for a detailed discussion, see][and references therein]{diesing23}. This assumption is not necessarily consistent with recent kinetic simulations \citep[e.g., ][]{zacharegkas+24}, rather, it represents a lower bound on the maximum proton energy, and thus our SNR's very high energy (VHE) $\gamma$-ray emission.

The propagation of CRs is expected to excite streaming instabilities, \citep[]{bell78a,bell04,amato+09,bykov+13}, which drive magnetic field amplification.
To approximate this amplification, we assume contributions from both the resonant streaming instability \citep[e.g., ][]{kulsrud+68,zweibel79,skilling75a, skilling75b, skilling75c, bell78a, lagage+83a}, and the non-resonant hybrid instability \citep{bell04}, using the recipe described in \cite{diesing+21, diesing23}, and in Paper I \citep[see also,][]{cristofari+21, zacharegkas+22}.
By the onset of the radiative phase, however, the relatively low shock velocity \citep[$\vsh < 600$ km s$^{-1}$, see][]{diesing+21} leads to weak magnetic field amplification driven only by the resonant instability, which saturates at $\delta B/B_0 \sim 1$. This mildly amplified magnetic field is not directly considered in our MHD simulations. As such, the true dynamical impact of magnetic fields in typical SNRs may be slightly larger than that considered in this work. In particular, in quasi-parallel regions (i.e., where particle acceleration and therefore magnetic field amplification are efficient), compression of the perpendicular component of the turbulent, amplified field may contribute to shell disruption.

This particle acceleration model yields the instantaneous spectrum of protons accelerated at each timestep of our MHD simulations. We convert this spectrum to an instantaneous electron spectrum using the analytical approximation calculated in \cite{zirakashvili+07}, setting the normalization of the electron spectrum relative to that of protons to be equal to $10^{-3}$, which yields good agreement with observations of Tycho's SNR \citep{morlino+12} and with first-principles shock studies \citep[e.g.,][]{park+15,gupta+24b}.

To account for energy losses--adiabatic, proton-proton (hadrons only), and synchrotron (electrons only)--we shift and weight our instantaneous spectra as in Paper I: each spectrum is treated as a shell of particles that expands according to the velocity profiles given by our MHD simulations. We then apply energy losses (and calculate nonthermal emission) based on the density and magnetic field at each shell's current location. Note that, as in Paper I, we find that the dominant contribution to the nonthermal emission produced after shell formation actually comes from populations of CRs accelerated at earlier times. In other words, our results are not sensitive to particle acceleration at late times, which may be inefficient. 

To calculate the spectral energy distribution (SED) of photons produced by our weighted particle distributions, we use the radiative processes code \texttt{naima} \citep[][]{naima}, which computes emission due to synchrotron, bremsstrahlung, inverse Compton (IC) and neutral pion decay processes for each shell of CRs. The target density for photon production, $n_{\rm H}(t)$, is taken to be the hydrogen number density at the current location of each shell. Meanwhile, the target magnetic field is taken to be the shell's initial amplified field immediately in front of the shock, compressed based on $n_{\rm H}(t)$ (assuming only transverse field components experience compression). 
While this field differs slightly from the field calculated by our MHD simulations, we find that the two are reasonably similar during the radiative phase. Finally we add together our shells of photons to generate a cumulative, multi-zone SED.

\section{Results} \label{sec:results}

\begin{figure}[ht]
    \centering
    \includegraphics[width=\linewidth, clip=true,trim= 35 0 20 25 ]{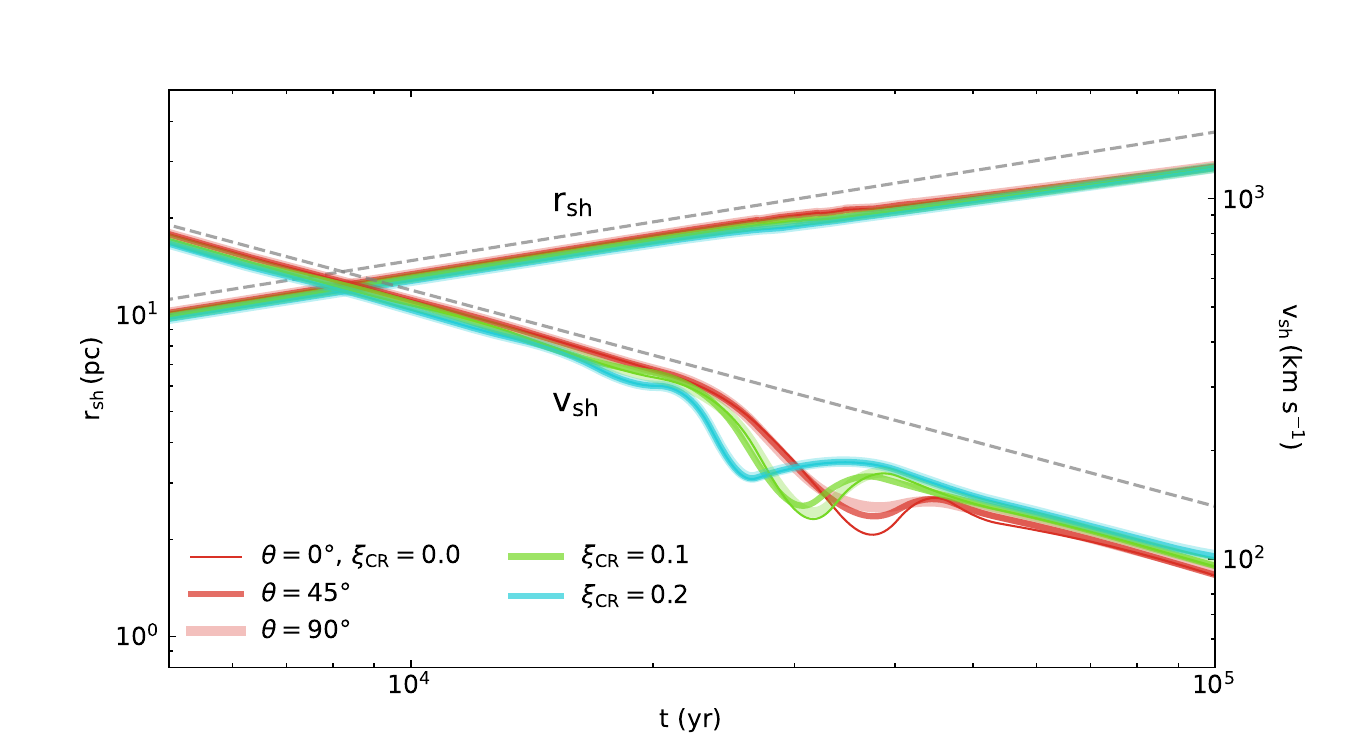}
    \caption{Shock radius ($r_{\rm sh}$) and velocity ($\vsh$) of our representative SNR ($E_{\rm SN} = 10^{51}$ erg, $M_{\rm ej} = 1 M_{\odot}$, $\nism = 1$ cm$^{-3}$, $B_{0} = 3 \mu G$) as a function of time. Gray dashed lines indicate the analytical Sedov-Taylor evolution, arbitrarily normalized (i.e., $r_{\rm sh} \propto t^{2/5}, \ \vsh \propto t^{-3/5}$). Line thickness denotes the inclination of the ambient magnetic field (taken to be 3 $\mu$G) with respect to the shock normal, $\theta$, while line color denotes the CR acceleration efficiency, $\xi_{\rm CR}$. Note that, in practice, we only treat the perpendicular component of the magnetic field, $B_{\perp} = B_0\sin{\theta}$, as the parallel component is not dynamically important and, in 1D spherical symmetry, violates $\nabla \cdot \mathbf{B} = 0$. Prior to the onset of the radiative phase, shock evolution is largely independent of CR or magnetic pressure. However, both play a role in mediating the timing of and shock deceleration associated with the onset of the radiative phase.}
    \label{fig:evolutions}
\end{figure}

\begin{figure}[ht]
    \centering
    \includegraphics[width=\linewidth, clip=true,trim= 40 60 80 90]{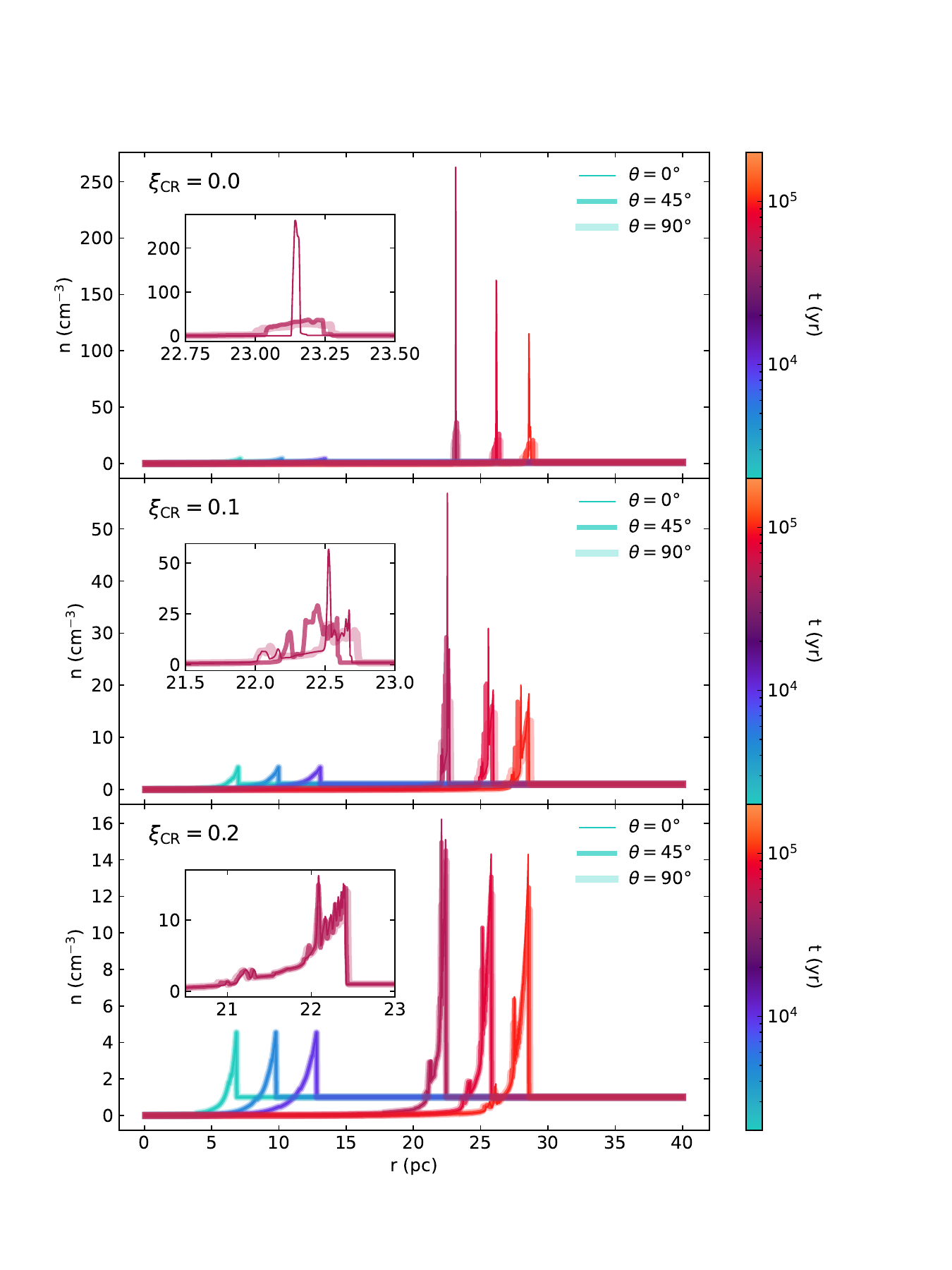}
    \caption{Density profiles of our representative SNR assuming $\xi_{\rm CR} = 0.0$ (top), $\xi_{\rm CR} = 0.1$ (middle), and $\xi_{\rm CR} = 0.2$ (bottom). The color scale denotes the age of the SNR, and the line thickness denotes magnetic field orientation with respect to the shock normal. The inset shows the region around the forward shock at $t = 5 \times 10^4$ yr. The presence of CRs and/or magnetic fields dramatically alters the density profile during the radiative stage, suppressing the maximum density and creating a complex shell structure.}
    \label{fig:profiles}
\end{figure}

In this section we present the results of our MHD simulations ( \ref{subsec:hydro_results}) and of our subsequent modeling of the nonthermal emission from our representative SNR ( \ref{subsec:emission_results}). Throughout this section, we consider acceleration efficiencies spanning $0.0 \leq \xi_{\rm CR} \leq 0.2$ and magnetic field inclinations (relative to the shock normal) spanning $0\degree \leq \theta \leq 90\degree$ by setting different $B_{\rm \perp}$.

\subsection{Shock hydrodynamics}
\label{subsec:hydro_results}
Shock evolutions ($r_{\rm sh}$ and $\vsh$ as a function of time) are shown in Figure \ref{fig:evolutions}, with line color denoting $\xi_{\rm CR}$ and line thickness denoting $\theta$. Here, the location of the shock is taken to be the outermost location where the compression ratio, $R \equiv \rho(r)/\rho_0$ exceeds 3.9. This shock-finding technique holds provided that the shock remains strongly supersonic, as is the case for the parameters and time period considered in Figure \ref{fig:evolutions}. The shock velocity is then calculated based on a spline derivative of $r_{\rm sh}$. Note that this calculation technique for $\vsh$ smooths out the characteristic oscillatory behavior of the forward shock velocity during the radiative phase \citep[e.g.,][]{petruk+18}. However, we find that changing between shock velocity estimation prescriptions has little bearing on the results shown in Section \ref{subsec:emission_results}, since the average shock velocity as a function of time remains roughly the same.

The onset of the radiative phase is clearly apparent in Figure \ref{fig:evolutions} as a dip in $\vsh$. Broadly speaking, the smaller $\xi_{\rm CR}$ and $\theta$, the deeper the dip. In other words, CR and magnetic pressure can mitigate the slowing of the shock at the onset of the radiative phase. 
This occurs because nonthermal pressures reduce the postshock temperature, allowing the cooling-efficient temperature, which is $\sim 10^5$ K, to be reached slightly earlier compared to the $\xi_{\rm CR}=0.0$ scenario.
However, beyond that initial dip, $\theta$ has very little impact on the overall shock evolution, since the magnetic pressure fraction is quite small except when highly compressed. CR pressure, on the other hand,  can constitute up to 20\% of the postshock pressure (by construction), meaning that it does not rely on strong compression to be dynamically relevant. Namely, shocks with large acceleration efficiencies expand somewhat more slowly and become radiative earlier, due to the higher compressibility of the shock (recall that $\gamma_{\rm CR} = 4/3$ whereas $\gamma_{\rm th} = 5/3$). However, they also recover a higher $\vsh$ during the radiative phase, since they act as a reservoir of pressure support that does not lose energy as the thermal gas does. This result is broadly consistent with that of \cite{diesing+18}.

Density profiles from our simulation suite are shown in Figure \ref{fig:profiles}, where each panel corresponds to a different $\xi_{\rm CR}$, line thickness once again represents $\theta$, and the color scale denotes the age of the shock. The onset of the radiative phase, which takes place at roughly $t = 2-4 \times 10^4$ yr, depending on $\xi_{\rm CR}$, is clearly apparent via the development of the characteristic dense shell behind the forward shock. As predicted, both CRs and magnetic fields provide pressure support that serves to mitigate the density of this shell. However, for high CR acceleration efficiency ($\xi_{\rm CR} = 0.2$; bottom panel of Figure \ref{fig:profiles}), changing $\theta$ has little to no effect on the resulting density profile. This can be understood as a sort of saturation; in the $\xi_{\rm CR} = 0.2$ case, the maximum achievable compression gives a magnetic pressure fraction that remains small relative to the $\xi_{\rm CR}$, meaning that changing $\theta$ does little to alter shock dynamics. We also note that the introduction of CR and magnetic pressures creates highly complex shell structures with respect to the case with thermal gas alone (see the insets in Figure \ref{fig:profiles}). This complexity can introduce variability in our nonthermal emission predictions (Section \ref{subsec:emission_results}), as CRs are advected into regions of higher or lower density.

\begin{figure}[ht]
    \centering
    \includegraphics[width=\linewidth, clip=true,trim= 30 60 55 90]{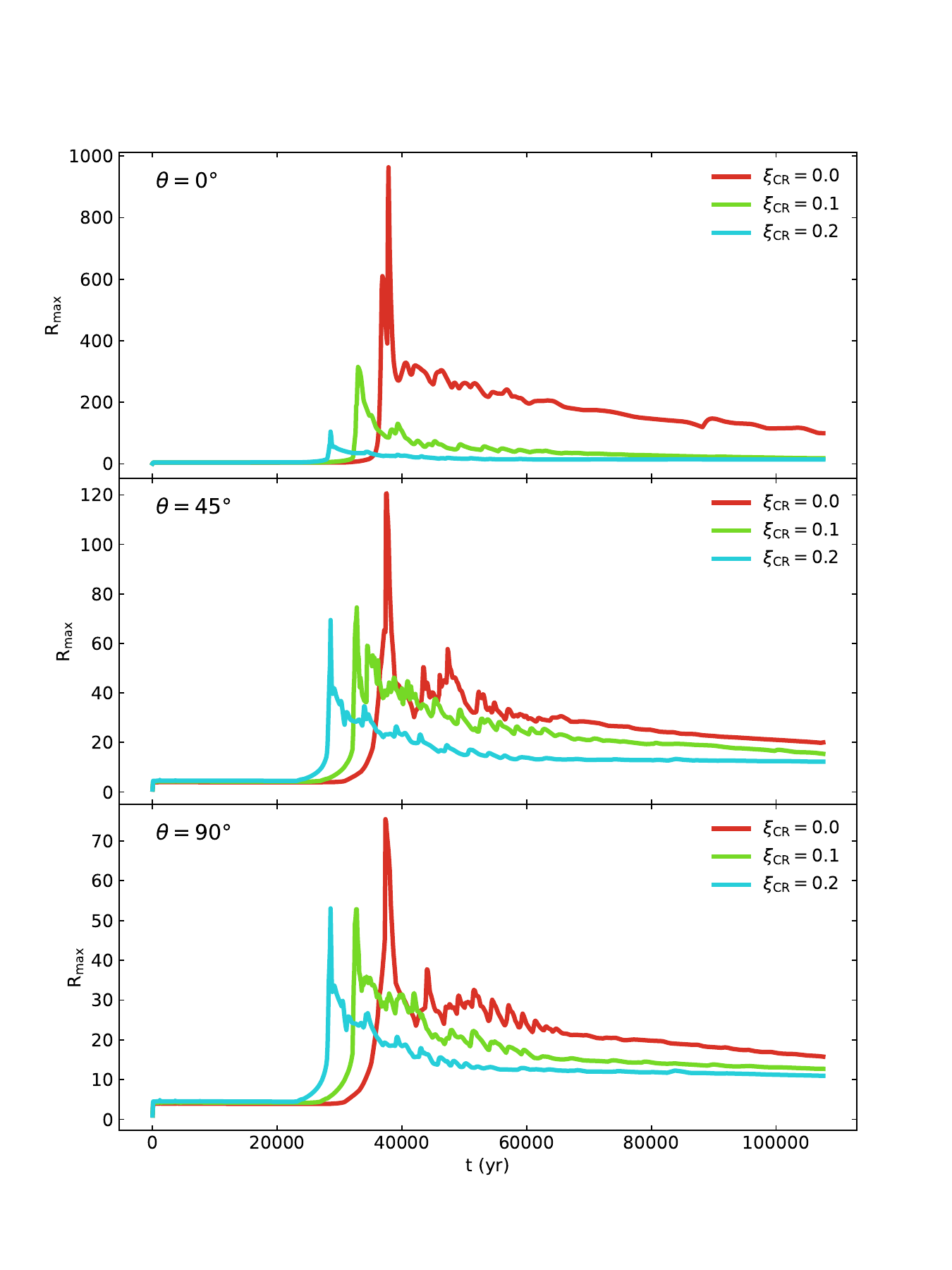}
    \caption{The maximum compression ratio, $R_{\rm max} \equiv \text{max}[\rho(r)/\rho_0]$ as a function of shock age for our representative SNR. Each panel corresponds to a different inclination, $\theta$, of the magnetic field (taken to be 3 $\mu$G) with respect to the shock normal, while line color denotes the CR acceleration efficiency, $\xi_{\rm CR}$. Under the right conditions, CRs and magnetic fields can individually decrease shell densities by up to an order of magnitude. }
    \label{fig:compression}
\end{figure}

To summarize the effect of CR and magnetic pressures on shell formation, we show the maximum compression ratio, $R_{\rm max} \equiv \text{max}[\rho(r)/\rho_0]$ as a function of shock age in Figure \ref{fig:compression}. This time, each panel corresponds to a different $\theta$ while color denotes $\xi_{\rm CR}$. Again, we see that increasing $\xi_{\rm CR}$ from 0.0 to 0.2 decreases $R_{\rm max}$ from $\sim 1000$ to $\sim 100$, a nearly tenfold reduction. Meanwhile, as one increases $\theta$ from $0 \degree$ to $90 \degree$ (assuming $\xi_{\rm CR} = 0.0$), $R_{\rm max}$ declines by a similar amount (from $\sim 1000$ to $\sim 80$). Thus, both CRs and magnetic fields can reduce the shell density. Keep in mind that we consider only the typical ISM value of $B_0 = 3 \mu G$; shell disruption could be even more dramatic in the presence of mild magnetic field amplification or a more highly magnetized medium. We also note that Figure \ref{fig:compression} shows that the effect of increasing $\xi_{\rm CR}$ decreases as $\theta$ increases. This effect can be described in terms similar to the saturation discussed above. Recall that, if one raises $\xi_{\rm CR}$, one lowers $R_{\rm max}$, thereby reducing the maximum magnetic pressure. Similarly, if one raises $\theta$, CRs experience less compression, and therefore provide less pressure support in the vicinity of the shell. In other words, the combined effect of CRs and magnetic fields is highly nonlinear. However, in practice, one would expect efficient CR acceleration only in quasi-parallel (small $\theta$) regions \citep[recall][]{caprioli+14a}. As such CRs and magnetic fields can provide reasonably even shell disruption across the surface of an SNR, with CRs providing pressure support where $\theta$ is small, and magnetic fields providing pressure support where $\theta$ is large.

\begin{figure}
    \centering
    \includegraphics[width=\linewidth, clip=true,trim= 40 5 80 30]{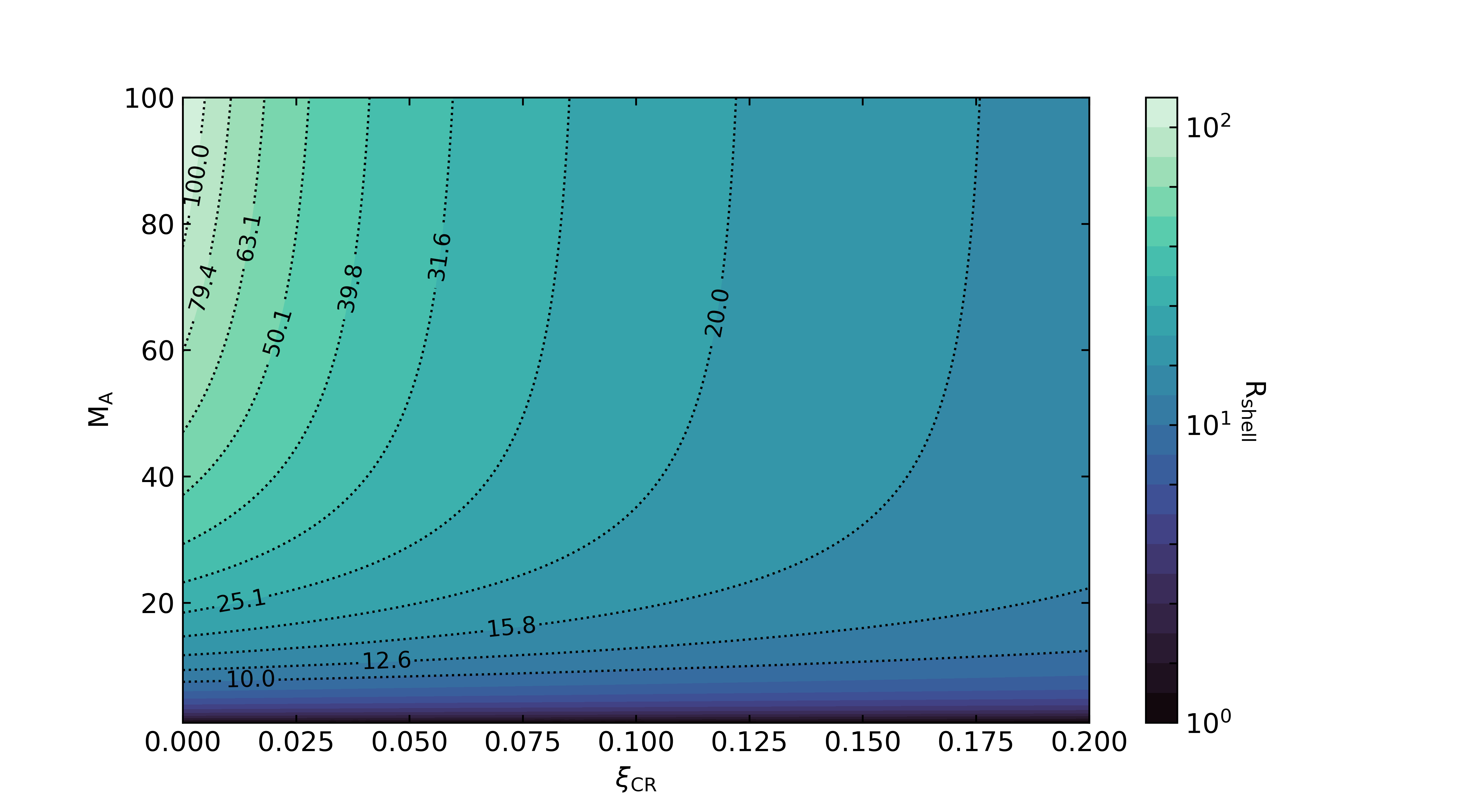}
    \caption{Predicted compression ratio of the radiative shell ($R_{\rm shell} \equiv \rho_{\rm shell}/\rho_0)$  a function of Alfv\'enic Mach number (considering only the perpendicular component of the magnetic field, $M_{\rm A} \equiv \vsh \sqrt{4\pi \rho_0}/ B_\perp$) and CR pressure fraction ($\xi_{\rm CR} \equiv P_{\rm CR}/(\rho_0 \vsh^2)$), assuming a sonic Mach number $M \simeq 21$ (i.e., the Mach number consistent with the ambient medium considered in our simulations, taking $\vsh \simeq 250$ km s$^{-1}$ at the onset of the radiative stage). Note that these predictions are based on a simple calculation that solves the modified shock-jump conditions (for a full derivation, see Appendix \ref{sec:jumpcond}), and are therefore approximate. That being said, this analytic formalism recovers the overall behavior exhibited in Figure \ref{fig:compression}, including the approximate values of $R_{\rm max}$.}
    \label{fig:compression_analytic}
\end{figure}

\begin{figure*}[ht]
    \subfloat{%
      \includegraphics[width=0.5\textwidth, clip=true,trim= 10 10 80 30]{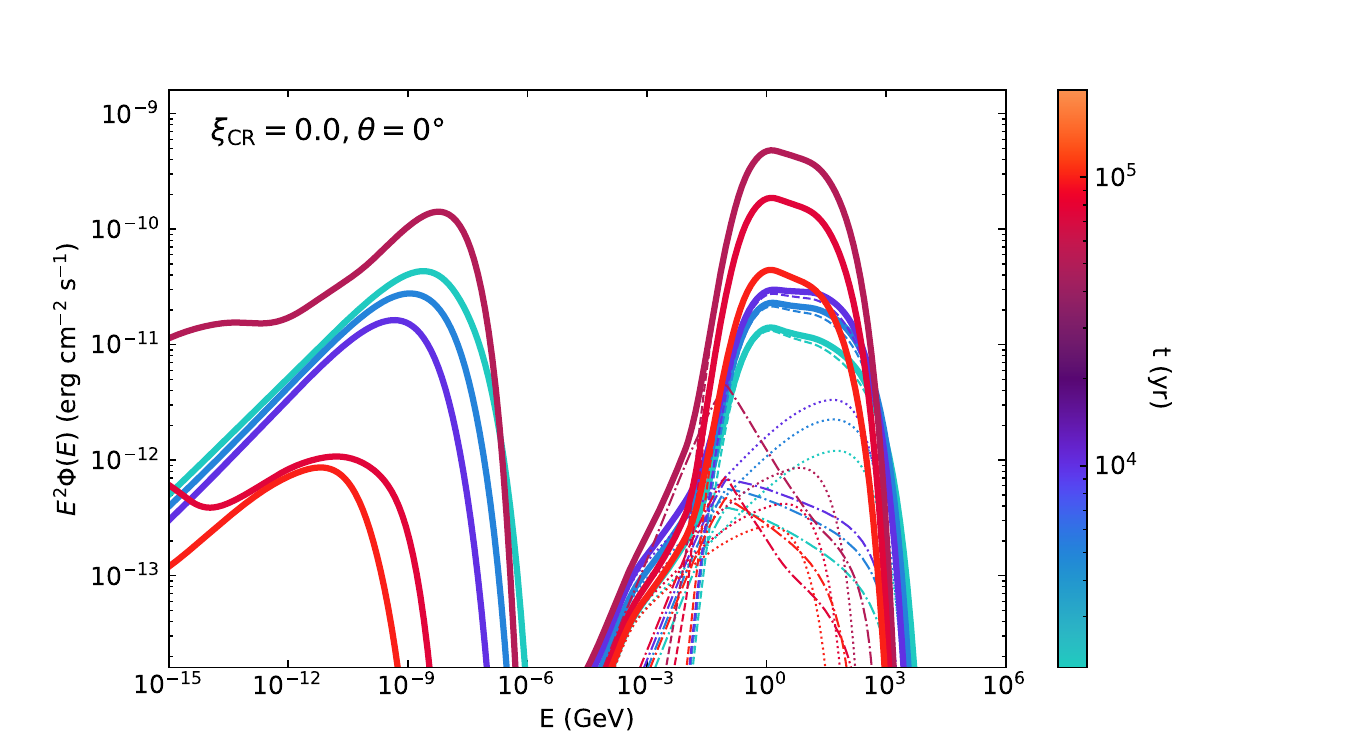}} 
    \subfloat{%
      \includegraphics[width=0.5\textwidth, clip=true,trim= 10 10 80 30]{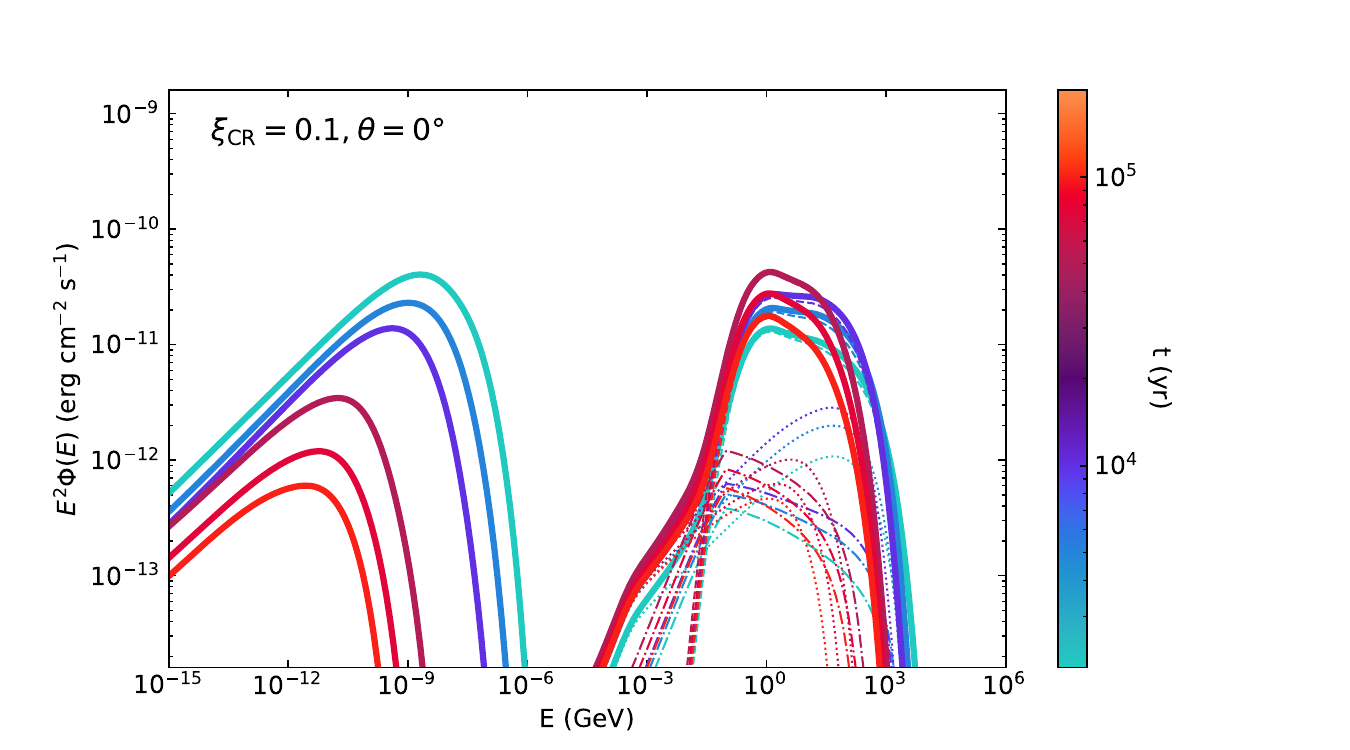}} \\
    \subfloat{%
      \includegraphics[width=0.5\textwidth, clip=true,trim= 10 10 80 30]{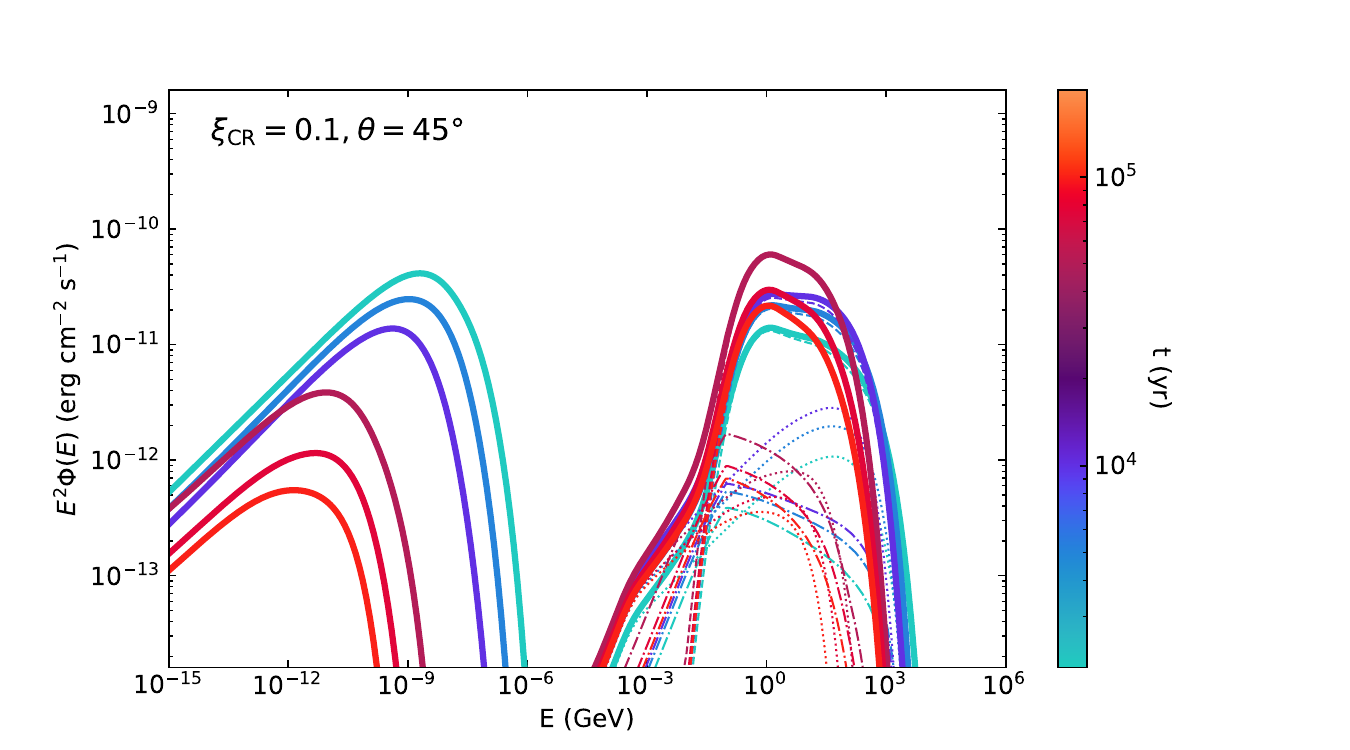}}
    \subfloat{%
      \includegraphics[width=0.5\textwidth, clip=true,trim= 10 10 80 30]{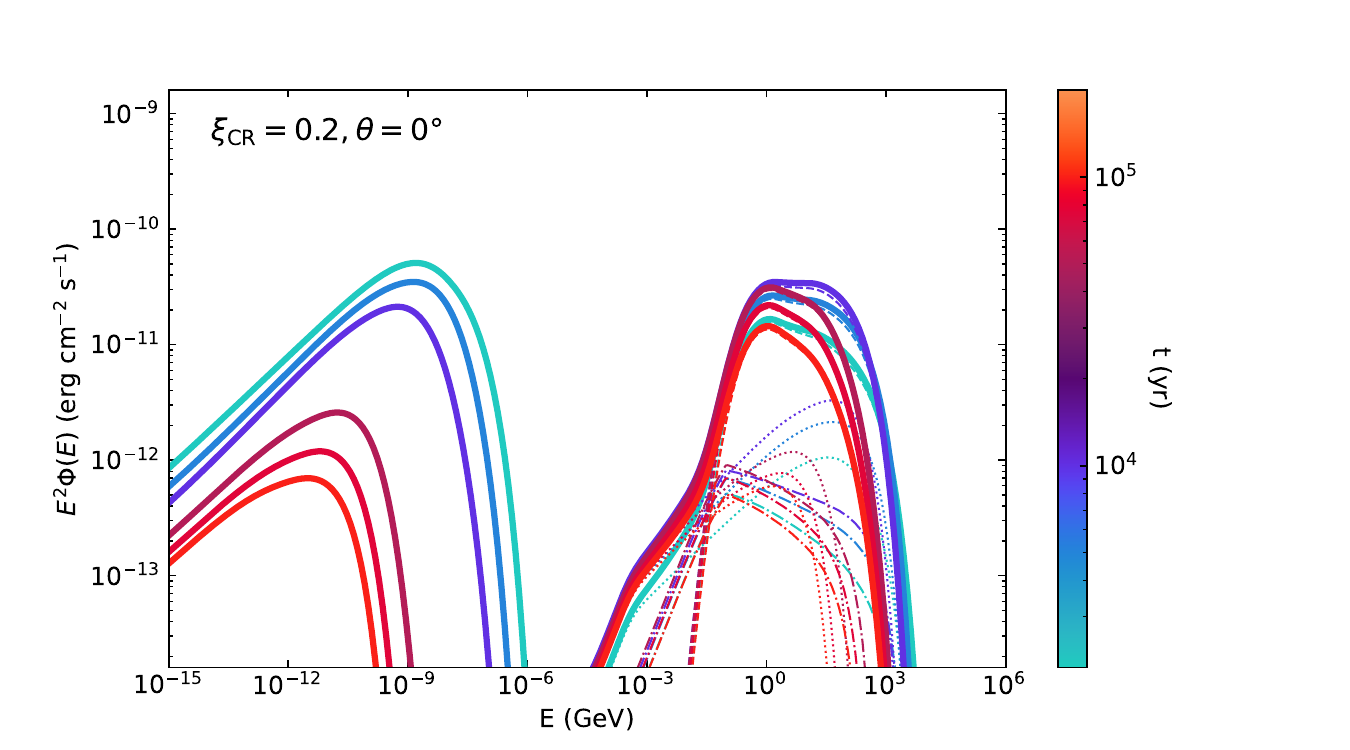}}

\caption{Nonthermal multi-wavelength SEDs from our representative SNR at a distance of 3 kpc. Each panel corresponds to a different magnetic field orientation and/or CR acceleration efficiency. The color scale denotes the age of the SNR, while line styles indicate different emission mechanisms: dashed corresponds to pion decay (hadronic emission), dot-dashed corresponds to nonthermal bremsstrahlung, and dotted corresponds to inverse Compton. Thick lines correspond to total emission, which is dominated by synchrotron at low energies (radio to X-rays) and pion decay at high energies ($\gamma$-rays).}
\label{fig:SEDs}
\end{figure*}

Note that the results shown in Figure \ref{fig:compression} can be explained (and extrapolated) using a simple, analytic formalism derived in Appendix \ref{sec:jumpcond}. Namely, by solving the shock jump conditions between the far upstream, immediate downstream (where the thermal gas has not yet cooled), and the radiative shell, we obtain the shell compression ratio ($R_{\rm shell} \equiv \rho_{\rm shell}/\rho_0$), for a given Alfv\'enic Mach number (considering only the perpendicular component of the magnetic field, $M_{\rm A} \equiv \vsh \sqrt{4\pi \rho_0}/ B_\perp$), sonic Mach number ($M \equiv \vsh/c_{\rm s}$), and CR pressure fraction ($\xi_{\rm CR} \equiv P_{\rm CR}/(\rho_0 \vsh^2)$). This is illustrated in Figure \ref{fig:compression_analytic} using a fixed $M \simeq 21$ (i.e., the Mach number consistent with the ambient medium considered in our simulations, assuming $\vsh \simeq 250$ km s$^{-1}$ at the onset of the radiative stage), while it can be extrapolated to different parameters. The shell density (compression ratios) obtained from this analysis are broadly consistent with those shown in Figure \ref{fig:compression}, and recover the nonlinear relationship between shell density, perpendicular magnetic field strength, and CR acceleration efficiency shown in Figure \ref{fig:compression}. 

\subsection{Observational impacts}
\label{subsec:emission_results}

While our simulations span parameter space in terms of CR acceleration efficiency ($\xi_{\rm CR}$) and magnetic field orientation ($\theta$), in this section we largely focus our attention on cases with medium to high $\xi_{\rm CR}$ and medium to low $\theta$: $\xi_{\rm CR} = 0.1, \theta = 0 \degree$; $\xi_{\rm CR} = 0.1, \theta = 45 \degree$; $\xi_{\rm CR} = 0.2, \theta = 0 \degree$. Namely, in the case of low $\xi_{\rm CR}$, we expect minimal nonthermal emission. Meanwhile, high $\xi_{\rm CR}$ can only be achieved when $\theta$ is relatively small. That being said, we do also consider the case with $\xi_{\rm CR} = 0.0$ and $\theta = 0\degree$ in our MHD simulations, assuming $\xi_{\rm CR} \simeq 0.1$ when calculating our nonthermal emission. This special case was discussed extensively in Paper I and is meant to act as a baseline for comparison purposes. Namely, it represents the scenario in which particles are accelerated in the usual way, but are not allowed to have a dynamical effect on the shock.

A sample of nonthermal spectral energy distributions (SEDs) from radio to $\gamma$-rays are shown in Figure \ref{fig:SEDs}, with each panel corresponding to a different $\xi_{\rm CR}$ and/or $\theta$. As in Figure \ref{fig:profiles}, the color scale denotes the age of the SNR. Clearly, the case with no dynamical impact from CRs or magnetic fields ($\xi_{\rm CR} = 0.0$ and $\theta = 0\degree$, top left panel of Figure \ref{fig:SEDs}) is unique, exhibiting a dramatic rise in both radio and $\gamma$-rays at the onset of the radiative phase. While the synchrotron emission falls off as the shock age approaches $10^5$ yr, largely due to strong synchrotron losses in the compressed amplified magnetic field (recall that we use this turbulent field as the target for synchrotron emission), the $\gamma$-ray emission remains enhanced out to $t = 10^5$ yr. Notably, this $\gamma$-ray emission is so large that TeV emission becomes non-negligible, despite the very conservative estimates of the maximum proton energy considered in this work. Namely, the rise is sufficient to generate appreciable emission above the high-energy cutoff. All of these results are broadly consistent with those shown in Paper I.
\begin{figure}[ht]
    \centering
    \includegraphics[width=\linewidth, clip=true,trim= 25 60 50 90]{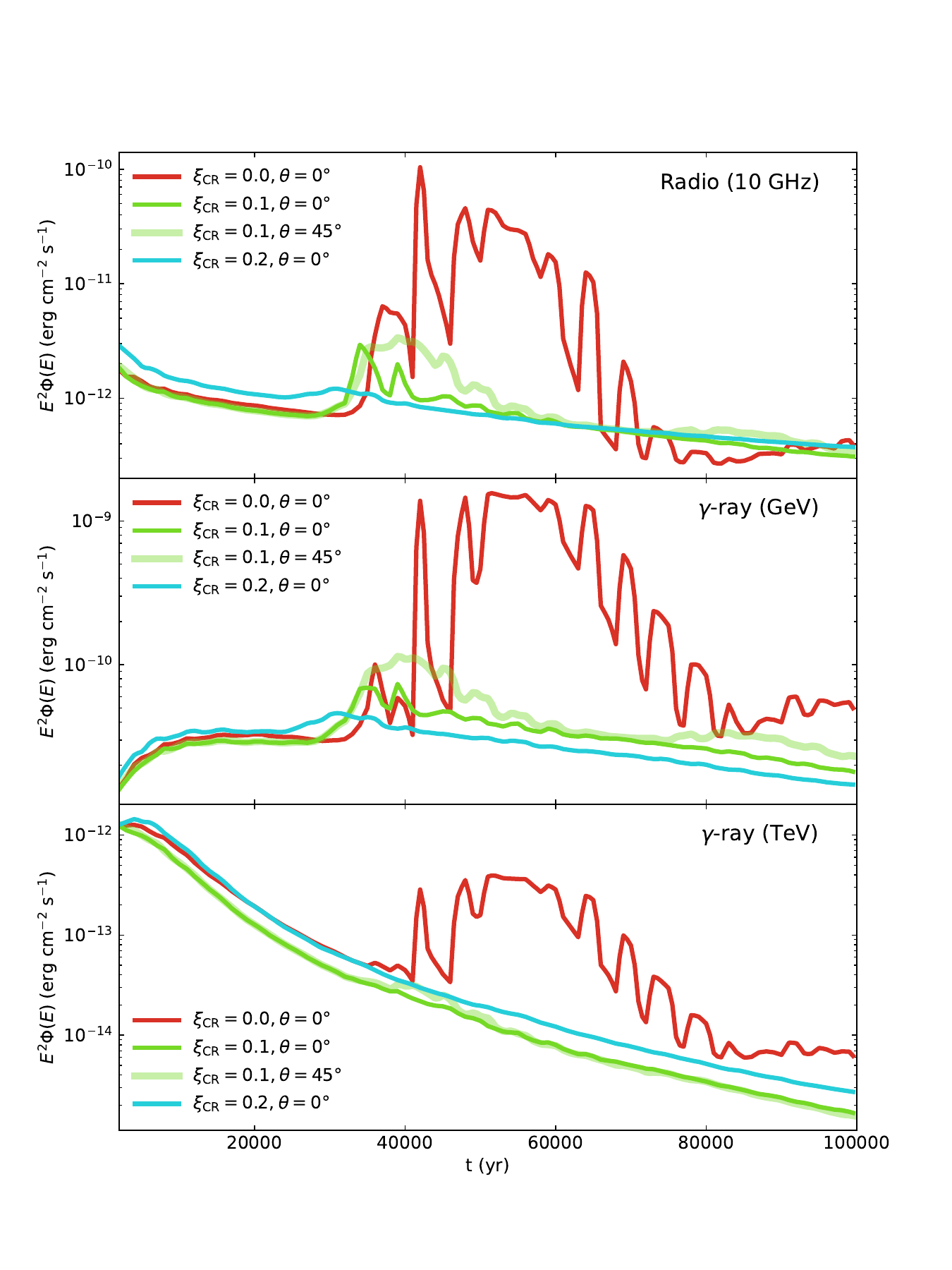}
    \caption{Nonthermal light curves for our representative SNR at a distance of 3 kpc. Each panel represents a different photon energy, while line thickness and color denote the magnetic field orientation and acceleration efficiency, respectively (similar to Figure \ref{fig:evolutions}). The presence of CRs reduces not only the magnitude of the nonthermal rebrightening at the onset of the radiative stage, but also its duration.}
    \label{fig:lightcurves}
\end{figure}
On the other hand, the presence of CRs mitigates this enhancement, with no appreciable rise in synchrotron emission and only a modest rise in the nonthermal $\gamma$-rays around the onset of the radiative phase; this $\gamma$-ray emission quickly decays to emission that resembles Sedov-Taylor levels. This impact can be seen more clearly in Figure \ref{fig:lightcurves}, which displays nonthermal light curves of our representative SNR in radio (top panel), GeV $\gamma$-rays (middle panel), and TeV $\gamma$-rays (bottom panel). 

In Figure \ref{fig:lightcurves}, colors represent different $\xi_{\rm CR}$ while line thickness denotes $\theta$, similar to Figures \ref{fig:evolutions} and \ref{fig:compression}. While the onset of the radiative phase does produce a rise in nonthermal emission even with CRs and compressed magnetic fields, this increase is strongly attenuated by the presence of nonthermal pressure. Meanwhile, at TeV energies, which probe particle energies above the high-energy cutoff, CRs and magnetic fields cause this rise to all but disappear. Notably, however, if nonthermal pressure is not important (i.e., the $\xi_{\rm CR} = 0.0, \ \theta = 0\degree$ case), the resultant rise in TeV emission marginally exceeds CTA detection limits \citep{CTA} for our representative SNR at $d = 3$ kpc, meaning that non-detections of TeV emission from nearby radiative SNRs such as IC443, W44, and Cygnus Loop (all of which have reported distances at or below 3 kpc) may definitively rule out commonly held assumptions regarding shell formation.

Intriguingly, closer inspection of Figure \ref{fig:lightcurves} reveals that the $\xi_{\rm CR} = 0.1, \ \theta = 45\degree$ case actually produces a longer-lasting rise in nonthermal emission than its quasi-parallel counterpart ($\xi_{\rm CR} = 0.1, \ \theta = 0\degree$). While the former case has a lower maximum shell density as shown in Figures \ref{fig:profiles} and \ref{fig:compression}, variations in the complex shape of the postshock density (and, equivalently, velocity) profile can cause particles to unexpectedly spend more or less time probing regions of high density (see the insets on Figure \ref{fig:profiles}). As such, we stress that our predictions can change slightly from one SNR to the next, especially in the presence of a nonuniform ambient medium. However, our overall result remains the same: CRs and magnetic fields dramatically reduce the emission we expect during the radiative phase of SNR evolution.
\begin{figure*}[ht]
    \subfloat{%
      \includegraphics[width=0.333\textwidth, clip=true,trim= 10 40 40 40]{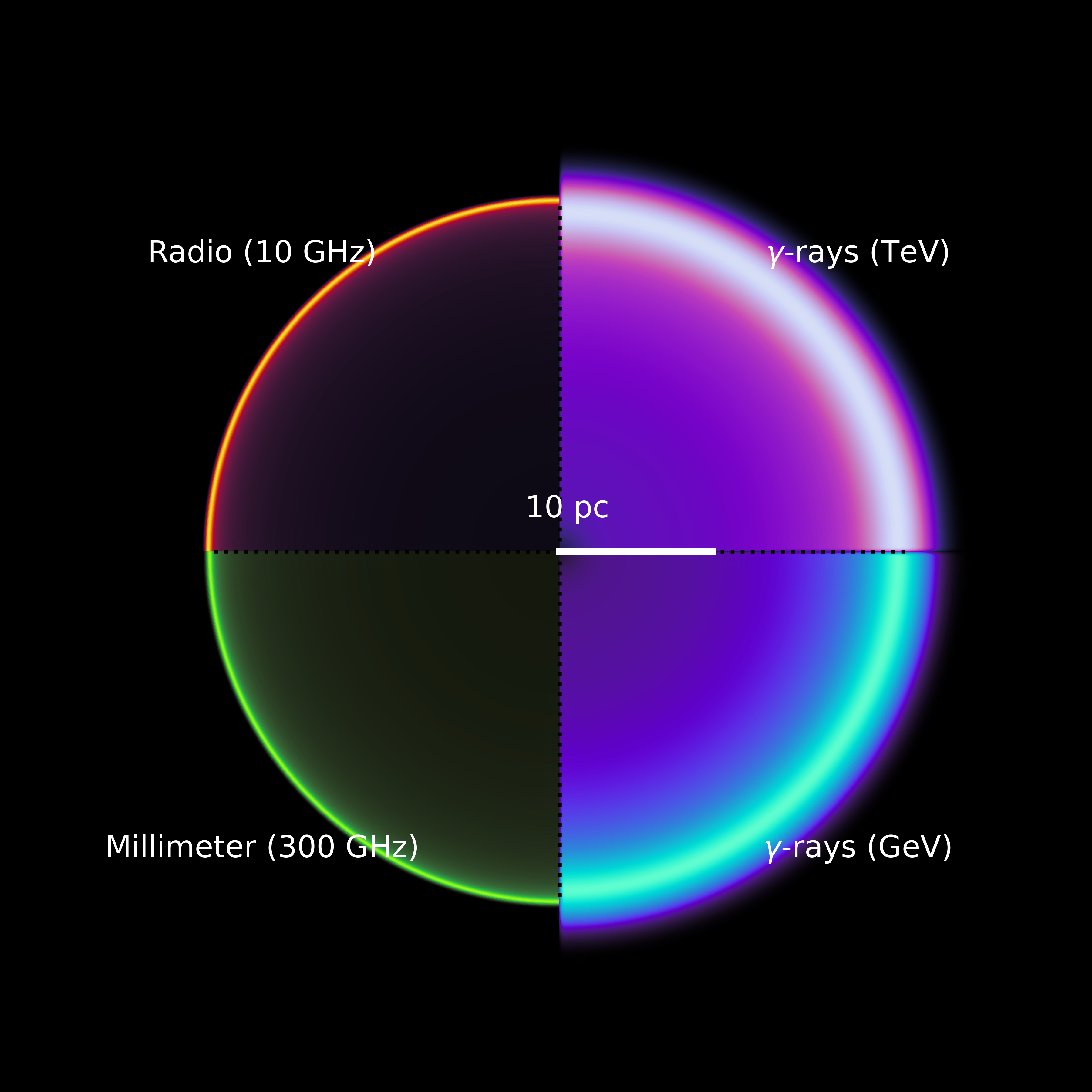}} 
    \subfloat{%
      \includegraphics[width=0.333\textwidth, clip=true,trim= 10 40 40 40]{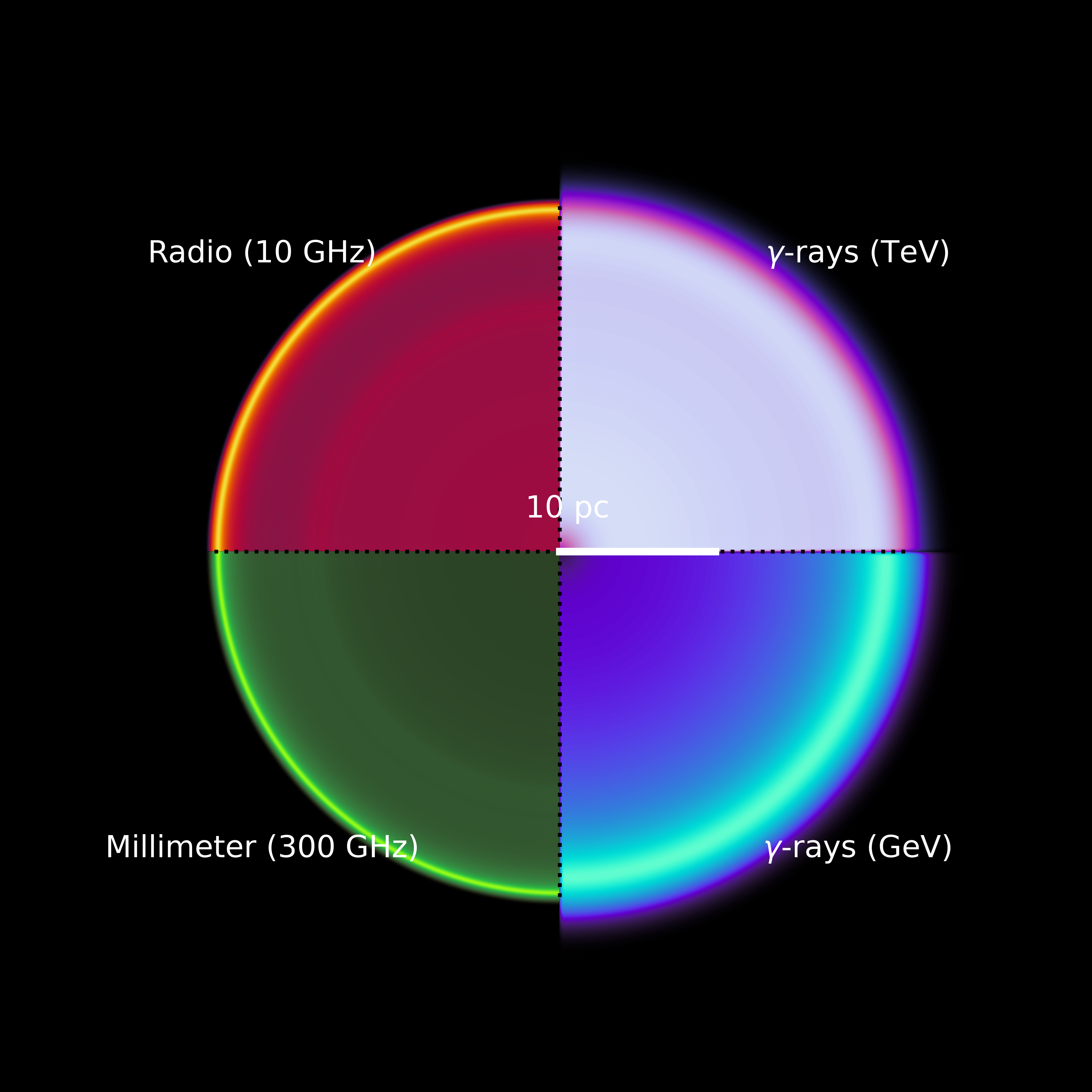}} 
    \subfloat{%
      \includegraphics[width=0.333\textwidth, clip=true,trim= 10 40 40 40]{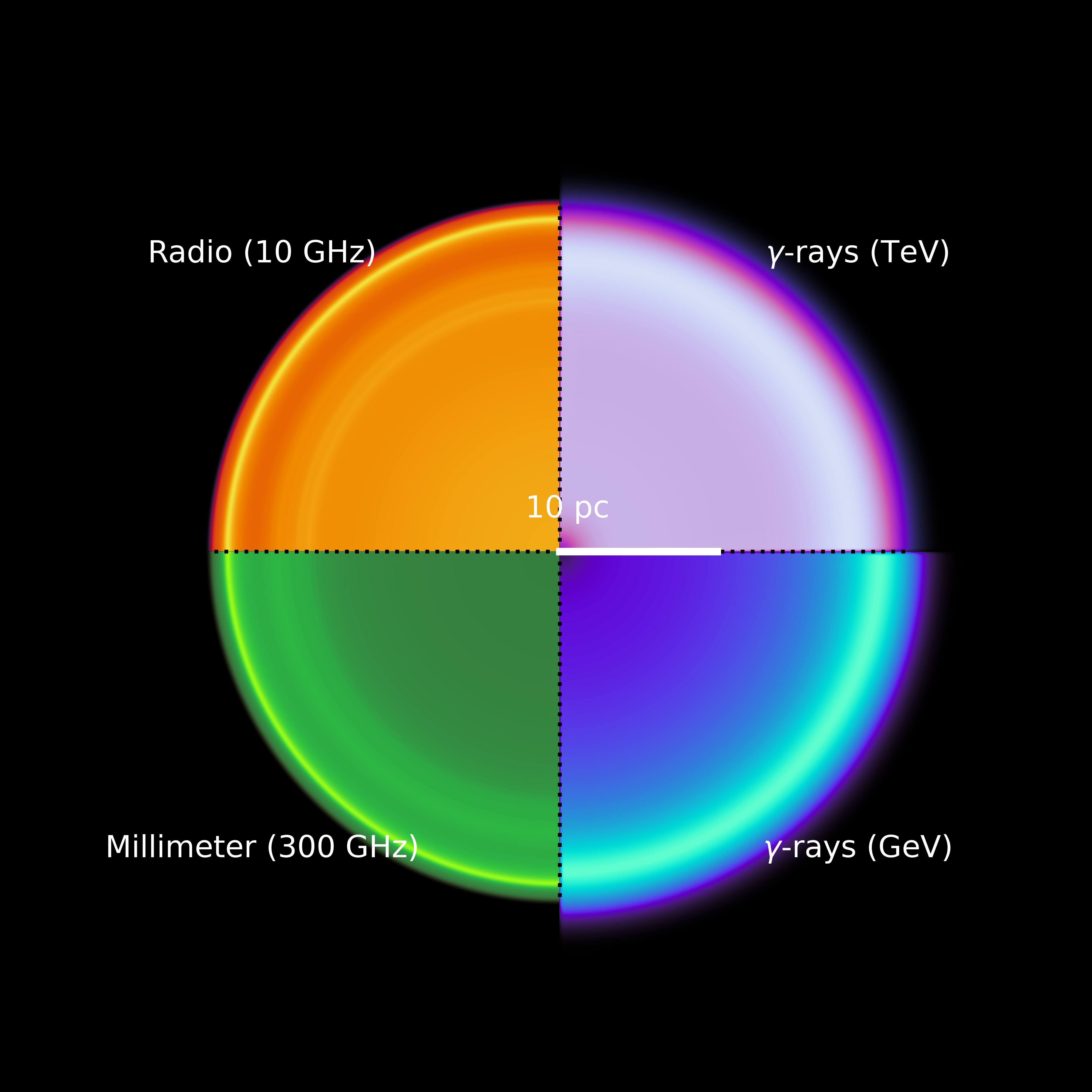}} 
    \caption{Mock nonthermal images of our representative SNR with $\theta  = 0\degree$ at $t = 5\times10^4$ yr, assuming $\xi_{\rm CR} = 0.0$ (left), $\xi_{\rm CR} = 0.1$ (middle), and $\xi_{\rm CR} = 0.2$ (right). Each quarter-image corresponds to a different photon energy with separate, arbitrary normalizations (in all cases, however, brightness scales with luminosity). The radio and millimeter maps have been convolved with a 2D Gaussian with $\sigma = 1.5\times10^{-1}$ pc, while the $\gamma$-ray maps have been convolved with a 2D Gaussian with $\sigma = 1.5$ pc, corresponding to the approximate resolution of the VLA and CTA, respectively, for nearby radiative SNR candidates such as IC443 or W44. While the dense shell that forms after the onset of the radiative phase is apparent regardless of acceleration efficiency, it is far more prominent if that efficiency is low. Moreover, at TeV energies, the shell only appears in the absence of CR pressure (which, in practice, would result little to no nonthermal emission).}
    \label{fig:image}
\end{figure*}

Finally, as in Paper I, we show mock, arbitrarily-normalized nonthermal images of our representative SNR at $t = 5 \times 10^4$ yr in Figure \ref{fig:image}. Each image quadrant corresponds to a different photon energy, while each panel corresponds to a different CR acceleration efficiency, assuming $\theta = 0.0$ (from left to right, $\xi_{\rm CR} = 0.0$, $\xi_{\rm CR} = 0.1$, and $\xi_{\rm CR} = 0.2$).  $\gamma$-ray images have been convolved with a Gaussian with standard deviation $\sigma = 1.5$ pc, in order to approximate the resolving capabilities of CTA \citep{CTA}, while radio and millimeter images are instead convolved with a Gaussian with $\sigma = 1.5\times10^{-1}$ pc in order to approximate the resolving capabilities of a telescope like the VLA \citep[e.g.,][]{castelletti+11}, assuming a nearby radiative SNR such as IC443 or W44. Of course, real evolved SNRs will not be nearly as spherical as the idealized case in shown Figure \ref{fig:image} but, based on density maps from realistic, 3D simulations (M. Guo et al. 2024, in prep), the shell structure, if present, will still be readily apparent.

While the shell appears in radio/millimeter regardless of $\xi_{\rm CR}$, the center of the SNR has a much higher relative brightness as $\xi_{\rm CR}$ increases. As such, for a real, aspherical SNR expanding into a less uniform medium, the shell of an SNR with high $\xi_{\rm CR}$ is far more likely to be lost among other, brighter signatures not considered here, such as molecular cloud interactions. Such confusion becomes even more likely when one considers the fact that $\xi_{\rm CR}$ depends on $\theta$ and therefore may vary across the surface of the forward shock.

Meanwhile, the most obvious impact of CR pressure on the nonthermal images shown in \ref{fig:image} appears at TeV energies. While GeV $\gamma$-rays are largely unaffected except in their normalization (recall that the images in Figure \ref{fig:image} are arbitrarily normalized), the TeV profile goes from a resolvable shell in the $\xi_{\rm CR} = 0.0, \ \theta = 0\degree$ case to a uniform distribution of negligible emission. Thus, if nonthermal pressure sources are not important or, equivalently, if shell formation proceeds as expected in the literature, we predict that CTA will be able to observe radiative shells. Conversely, non-detection of these shells represents strong evidence for a meaningful impact of nonthermal pressure sources on SNR evolution. Note that instabilities related to the thermal gas \citep[e.g., pressure-driven thin shell overstability and nonlinear thin-shell instability, see][]{vishniac83, blondin+98} are \emph{not} expected to cause meaningful shell disruption on the timescales considered in this work \citep[e.g.,][]{kim+15}. Moreover, dense shells are clearly seen in 3D hydrodynamic simulations of radiative SNRs expanding into highly nonuniform media (M. Guo et al. 2024, in prep).

\section{Discussion}\label{sec:discussion}

In this section, we briefly examine our results in the context of SNR feedback. Semi-analytic work based on the thin-shell approximation--in which the swept-up mass is assumed to be confined to a thin shell behind the shock and the SNR expands due to pressure inside the hot bubble--suggests that the presence of CRs can extend the lifetimes of SNRs, enhancing the momentum they deposit in their host galaxy by a factor of $\sim 2-3$ for $\xi_{\rm CR} \simeq 0.1$ \citep[][]{diesing+18}. Namely, after the SNR enters the radiative phase and the thermal gas loses its energy, CRs can serve as an additional pressure source that sustains SNR expansion. In broad terms, MHD simulations performed in \cite{montero+22} corroborate this result, albeit with a much smaller enhancement of momentum deposition. However, these simulations include CR diffusion with a coefficient that is only mildly suppressed with respect to that of the Galaxy at large \citep[in practice, one expects strong suppression of the diffusion coefficient in the vicinity of an SNR][]{caprioli+14c}, meaning that these simulations likely underestimate the impact of CRs by overestimating their escape.

In contrast, we estimate the momentum deposition from our representative SNR with different acceleration efficiencies assuming CR transport is primarily advective, a reasonable approximation for the well-confined GeV particles responsible for the majority of the CR pressure. Since magnetic fields have little to no impact on the overall evolution of the shock (magnetic pressure is negligible except at the initial onset of the radiative phase when magnetic fields are compressed, see Figure \ref{fig:evolutions}), we consider only the $\theta = 0\degree$ case and run it with different $\xi_{\rm CR}$ until the average pressure inside the SNR equilibrates with that of the ISM. Note that this pressure equilibration condition corresponds to the total momentum becoming constant, since $dp/dt = 4\pi r^2(P_{\rm SNR}-P_{\rm ISM})$, where $p$ is the momentum and $P_{\rm SNR}$ and $P_{\rm ISM}$ are the total pressures inside and outside of the SNR, respectively. We then calculate the momentum of the SNR as a function of time, 
\begin{equation}
    p = \int_{0}^{r_{\rm sh}} 4 \pi r^2 \rho(r) v(r) dr,
\end{equation}
taking the total momentum deposited into the ISM to be the momentum at pressure equilibration.

\begin{figure}[ht]
    \centering
    \includegraphics[width=\linewidth, clip=true,trim= 35 0 50 25 ]{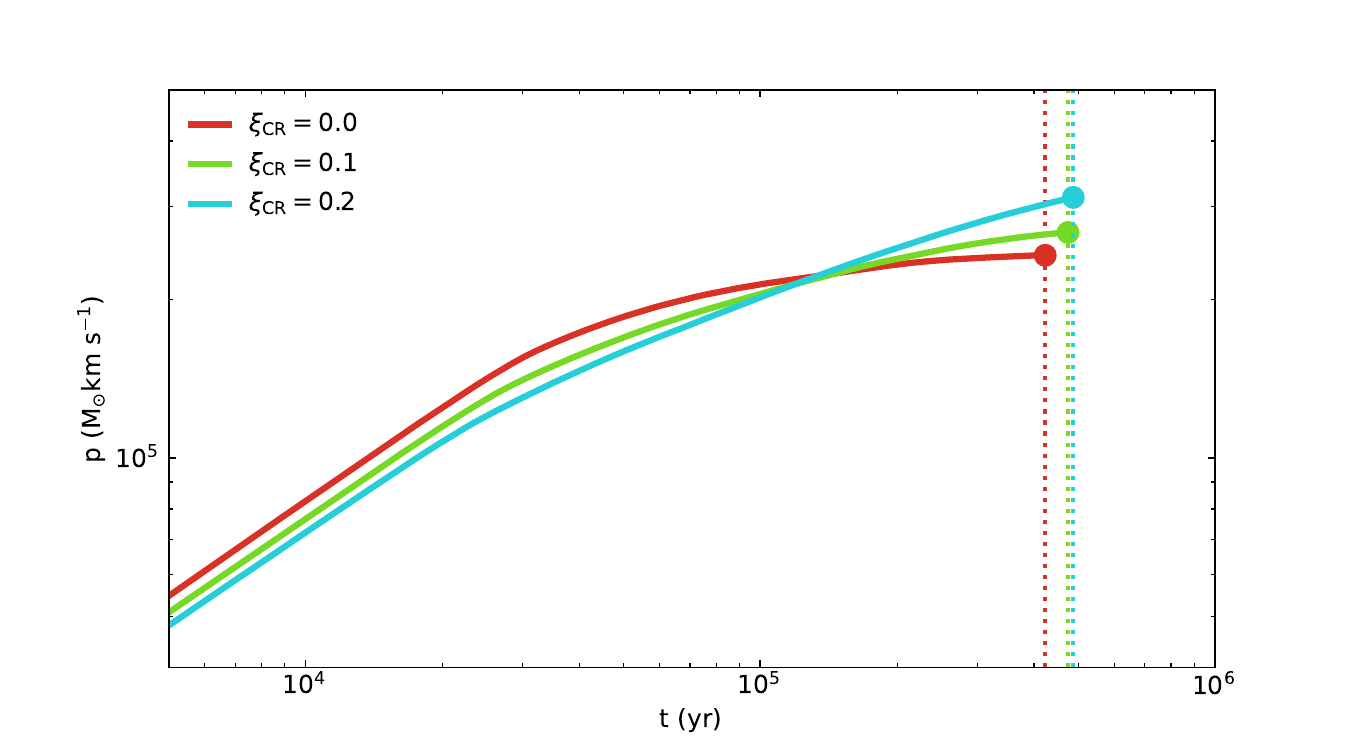}
    \caption{Momentum as a function of time for our representative SNR with different CR acceleration efficiencies, $\xi_{\rm CR}$, denoted by line color. The end of SNR evolution is approximated as pressure equilibration between the SNR and the ISM, and is denoted with dotted vertical lines. As in \cite{diesing+18}, efficient CR acceleration extends the life of a typical SNR and enhances the momentum it deposits into the ISM.}
    \label{fig:momentum}
\end{figure}

Momentum as a function of time for different $\xi_{\rm CR}$ is shown in Figure \ref{fig:momentum}. For the $\xi_{\rm CR} = 0.0$ case, we obtain a final momentum $p_{\rm f} \simeq 2.4 \times 10^5 \ M_{\odot} {\rm \ km \ s^{-1}}$, roughly consistent with the results of \cite{kim+15}. In the case of efficient CR acceleration, we obtain $p_{\rm f} \simeq 2.7 \times 10^5 \ M_{\odot} {\rm \ km \ s^{-1}}$ and $p_{\rm f} \simeq 3.1 \times 10^5 \ M_{\odot} {\rm \ km \ s^{-1}}$ for $\xi_{\rm CR} = 0.1$ and $\xi_{\rm CR} = 0.2$, respectively. This modest increase in momentum deposition is much smaller than the factor of 2-3 obtained in \cite{diesing+18}. We attribute this difference to the fact that CRs must do work in order to prevent collapse of the radiative shell, an effect that is not accounted for with the thin-shell approximation employed in \cite{diesing+18}.
Note that, while we only consider the $\theta = 0$ case in Figure \ref{fig:momentum}, changing the magnetic field does not alter our result; the magnetic field is 
dynamically important when highly compressed (i.e., in the region of the shell), meaning that it does not overall shock evolution.

\section{Conclusion} \label{sec:conclusion}

In summary, we modeled a representative SNR in the presence of CRs and magnetic fields in order to quantify their dynamical effect on SNR evolution. We find that both CRs and magnetic fields, while unimportant during the ejecta-dominated and Sedov-Taylor phases, can meaningfully alter SNR dynamics after the shock becomes radiative. Namely, both pressures serve to disrupt the collapse of the cold, dense shell behind the shock. More quantitatively, increasing the acceleration efficiency from $\xi_{\rm CR} = 0.0$ to $\xi_{\rm CR} = 0.1$ reduces shell densities by a factor of $\sim 3$, while rotating a $3\mu G$ ambient magnetic field from $\theta=0\degree$ to $\theta=90\degree$ with respect to the shock normal reduces shell densities by a factor of $\sim 10$. Simultaneously increasing $\xi_{\rm CR}$ and $\theta$ does not have a cumulative effect on shell density, since the available nonthermal pressure in the vicinity of the shell depends on compression. Moreover, in practice, an increase in $\theta$ is expected to decrease $\xi_{\rm CR}$, since CR injection declines as the inclination of the ambient magnetic field increases \citep[particularly for $\theta \gtrsim 50\degree$, see][]{caprioli+14a}.

Shell disruption due to CR and magnetic pressures has a dramatic impact on the nonthermal emission expected from a typical SNR. Namely, the enhanced densities and magnetic fields associated with radiative shells represent excellent targets for proton-proton and synchrotron emission, respectively. As such, in the absence of nonthermal pressure sources, the onset of the radiative phase ought to be one of the most luminous phases of SNR evolution, as discussed in Paper I. In contrast, when CR and magnetic pressures are included, this brightening during the radiative stage all but disappears, yielding nonthermal emission predictions that are more consistent with observations. Notably, CTA will be able to definitively test whether shells are disrupted, since the presence of a typical shell (i.e., without CRs or magnetic fields) yields bright ring of emission even at TeV energies.

CRs can also affect SNR feedback as in \cite{diesing+18} by extending the life of the SNR and enhancing the momentum it deposits into the ISM. However, this effect is quite small due to the fact that CRs lose energy as they do the work required to disrupt the formation of the radiative shell, and due to the fact that CR acceleration declines at late times.

In short, CRs and magnetic fields can meaningfully alter SNR evolution and emission. In particular, the dearth of radiative SNRs exhibiting bright, nonthermal emission with complete, shell-like morphologies represents strong evidence for shell disruption, likely by the nonthermal pressure sources described in this work. 

\acknowledgements 
We thank Chang-Goo Kim for useful comments and discussion. RD gratefully acknowledges support from the Institute for Advanced Study's Fund for Natural Sciences and the Ralph E. and Doris M. Hansmann Member Fund. SG acknowledges financial support from Princeton University.

\begin{appendix}

\section{Estimating the Density of the Radiative Shell}\label{sec:jumpcond}

Herein we describe the analytic framework we use to approximate the compression ratio of the shell, $R_{\rm shell}$, for an arbitrary SNR, as illustrated in Figure \ref{fig:compression_analytic}. We calculate $R_{\rm shell}$--including the effects of CRs and magnetic fields--by solving the equations for conservation of mass, momentum, and energy across the shock in a manner similar to that described in \cite{diesing+21}, \cite{haggerty+20} and \citep[]{gupta+21a}. For a 1D, stationary shock, these equations read,

\begin{equation}
    \rho(x)u(x) = \rho_1v_1,
\end{equation}
\begin{equation}
     \rho(x)v(x)^2 + P_{\rm th}(x) + P_{\rm CR}(x) + P_{\rm B}(x) = \rho_1v_1^2 + P_{\rm th,1} + P_{\rm B,1}, \text{ and}
     \label{eq:pcons}
\end{equation}
\begin{equation}
    \frac{\rho(x)v^3(x)}{2} + F_{\rm th}(x) + F_{\rm CR}(x) + F_{\rm B}(x) =  \frac{\rho_1v_1^3}{2} + F_{\rm th,1} + F_{\rm B,1},
    \label{eq:econs}
\end{equation}
where $v$ refers to the fluid velocity, $F$ refers to the enthalpy flux, and subscripts th, CR, and B refer to the gas, CR and magnetic field components respectively. Note that Equation \ref{eq:econs} is only valid from the far upstream to the region immediately behind the shock, as energy is not conserved in the radiative shell. To solve for $R_{\rm shell}$, we consider three regions, denoted with subscripts 1, 2, and 3:

\begin{itemize}
    \item Region 1: the region far upstream of the shock, i.e., the ambient ISM. We take the CR pressure and enthalpy flux to be equal to zero in this region (i.e., we neglect the CR escape flux).
    \item Region 2: the region immediately downstream of the shock, in which the ISM is shock-heated and has not had sufficient time to cool. As such, the compression ratio between Regions 1 and 2, denoted $R_{12} \equiv \rho_2/\rho_1$, can be estimated assuming conservation of energy (i.e., using Equation \ref{eq:econs}).
    \item Region 3: the radiative shell, located approximately one cooling length behind the shock. 
    Assuming efficient cooling reduces the shell temperature $T_{3}$ to the far-upstream temperature $T_{1}$ (or the cooling-heating equilibrium temperature), we approximate $P_{\rm th, 3}\approx \rho_{\rm 3} c_{\rm s,3}^2\equiv R_{\rm 13} (T_{\rm 3}/T_{\rm 1}) /(\gamma_{\rm th}M^2)$.
    Our eventual goal is to use conservation of momentum (i.e., Equation \ref{eq:pcons}) to find the compression ratio between this region and Region 1 (i.e., $R_{13} \equiv R_{\rm shell} \equiv \rho_3/\rho_1$).
\end{itemize}

We also make a number of simplifying assumptions.
Namely, we only consider the pressure and energy flux due to the perpendicular component of the magnetic field, $B_\perp$, as parallel components are not compressed and therefore do not have a strong dynamical impact. We also neglect magnetic field amplification (which is minimal during the radiative stage), as well as the kinetic energy of any associated plasma fluctuations. Together, these assumptions allow us to approximate $B \propto \rho$ or, equivalently, to set $P_{\rm B}(x) = P_{\rm B,1}R^2(x)$, where $R(x) \equiv \rho(x)/\rho_1$. Furthermore, we neglect any drifts of CRs with respect to the background plasma (i.e., the \emph{postcursor} described in \cite{diesing+21}), as these are negligible when magnetic field amplification is inefficient.

In Regions 1 and 2, the gas, CR, and magnetic energy fluxes can be written as,
\begin{equation}
    F_{\rm th}(x) = \frac{\gth}{\gth-1}v(x)P_{\rm th}(x), 
    \label{eq:theos}
\end{equation}
\begin{equation}
    F_{\rm CR}(x) = \frac{\gcr}{\gcr-1}v(x)P_{\rm CR}(x), \text{ and}
    \label{eq:CReos}
\end{equation}
\begin{equation}
    F_{\rm B}(x) = 2v(x)P_{\rm B}(x),
    \label{eq:Beos}
\end{equation}
where $\gth = 5/3$ is the adiabatic index of the thermal gas and $\gcr = 4/3$ is the adiabatic index of the CRs. 

We first obtain $R_{12}$ by solving Equations \ref{eq:pcons} and \ref{eq:econs} from Region 1 to Region 2. By dividing by $\rho_1 v_1^2$, Equation \ref{eq:pcons} can be rewritten as,
\begin{equation}
    1 + \frac{1}{\gth M^2} + \frac{1}{2 M_{\rm A}^2} = \frac{1}{R_{12}} + \xi_{\rm th,2} + \xi_{\rm CR} + \frac{R_{12}^2}{2 M_{\rm A}^2},
    \label{eq:pcons_norm1}
\end{equation}
where $\xi_{\rm x} \equiv P_{\rm x}/(\rho_1 v_1^2)$. Meanwhile, substituting Equations \ref{eq:theos}, \ref{eq:CReos}, and \ref{eq:Beos} into Equation \ref{eq:econs} and dividing by $\rho_1v_1^3/2$, we find,
\begin{equation}
    1 + \frac{\eta_{\rm th}}{\gth M^2} + \frac{2}{M_{\rm A}^2} = \frac{1}{R_{12}^2} + \frac{\eta_{\rm th}\xi_{\rm th,2}}{R_{12}} + \frac{\eta_{\rm CR}\xi_{\rm CR}}{R_{12}} + \frac{2R_{12}}{M_{\rm A}^2},
    \label{eq:econs_norm}
\end{equation}
where $\eta_{\rm x} \equiv 2\gamma_{\rm x}/(\gamma_{\rm x}-1)$. Thus, for a given $M$, $M_{\rm A}$, and $\xi_{\rm CR}$, Equations \ref{eq:econs_norm} and \ref{eq:pcons_norm1} can be solved for the unknown values of $\xi_{\rm th,2}$ and, more importantly, $R_{12}$. From here, we can solve Equation \ref{eq:pcons} from Region 1 to Region 3. Again, we divide by $\rho_1 v_1^2$ to obtain,
\begin{equation}
 1 + \frac{1}{\gth M^2} + \frac{1}{2 M_{\rm A}^2} = \frac{1}{R_{13}} + \frac{R_{\rm 13}}{\gamma_{\rm th}M^2} + \xi_{\rm CR}\bigg(\frac{R_{13}}{R_{12}}\bigg)^{\gcr} + \frac{R_{13}^2}{2 M_{\rm A}^2},
    \label{eq:pcons_norm2}
\end{equation}
where we assume that the gas temperature $T_{\rm 3} \simeq T_{\rm 1}$ and CRs are compressed adiabatically between Region 2 and Region 3.
Using the value of $R_{12}$ we obtained previously, it is straightforward to solve Equation \ref{eq:pcons_norm2} numerically to obtain $R_{13} = R_{\rm shell}$.
Note that in the absence of CRs and magnetic field, we recover $R_{\rm 13}\approx \gamma_{\rm th} M^2$, as is traditionally assumed for radiative shocks.
\end{appendix}
\end{CJK*}

\bibliographystyle{aasjournal}

\begin{thebibliography}{}
\expandafter\ifx\csname natexlab\endcsname\relax\def\natexlab#1{#1}\fi
\providecommand{\url}[1]{\href{#1}{#1}}
\providecommand{\dodoi}[1]{doi:~\href{http://doi.org/#1}{\nolinkurl{#1}}}
\providecommand{\doeprint}[1]{\href{http://ascl.net/#1}{\nolinkurl{http://ascl.net/#1}}}
\providecommand{\doarXiv}[1]{\href{https://arxiv.org/abs/#1}{\nolinkurl{https://arxiv.org/abs/#1}}}

\bibitem[{{Ackermann et al.}(2013)}]{ackermann+13}
{Ackermann et al.}, M. 2013, Science, 339, 807, \dodoi{10.1126/science.1231160}

\bibitem[{{Actis} {et~al.}(2011){Actis}, {Agnetta}, {Aharonian}, {Akhperjanian}, {Aleksi{\'c}}, {Aliu}, {Allan}, {Allekotte}, {Antico}, {Antonelli}, \& et~al.}]{CTA}
{Actis}, M., {Agnetta}, G., {Aharonian}, F., {et~al.} 2011, Experimental Astronomy, 32, 193, \dodoi{10.1007/s10686-011-9247-0}

\bibitem[{Agertz {et~al.}(2013)Agertz, Kravtsov, Leitner, \& Gnedin}]{agertz+13}
Agertz, O., Kravtsov, A.~V., Leitner, S.~N., \& Gnedin, N.~Y. 2013, \apj, 770, 25, \dodoi{10.1088/0004-637X/770/1/25}

\bibitem[{{Amato} \& {Blasi}(2005)}]{amato+05}
{Amato}, E., \& {Blasi}, P. 2005, MNRAS, 364, L76, \dodoi{10.1111/j.1745-3933.2005.00110.x}

\bibitem[{{Amato} \& {Blasi}(2006)}]{amato+06}
---. 2006, MNRAS, 371, 1251, \dodoi{10.1111/j.1365-2966.2006.10739.x}

\bibitem[{{Amato} \& {Blasi}(2009)}]{amato+09}
---. 2009, MNRAS, 392, 1591, \dodoi{10.1111/j.1365-2966.2008.14200.x}

\bibitem[{{Axford} {et~al.}(1977){Axford}, {Leer}, \& {Skadron}}]{axford+77p}
{Axford}, W.~I., {Leer}, E., \& {Skadron}, G. 1977, in International Cosmic Ray Conference, Vol.~2, \emph{Acceleration of Cosmic Rays at Shock Fronts}, 273--+.
\newblock \url{http://adsabs.harvard.edu/abs/1977ICRC....2..273A}

\bibitem[{{Bandiera} \& {Petruk}(2010)}]{bandiera+10}
{Bandiera}, R., \& {Petruk}, O. 2010, \aap, 509, A34, \dodoi{10.1051/0004-6361/200912244}

\bibitem[{{Bell}(1978)}]{bell78a}
{Bell}, A.~R. 1978, MNRAS, 182, 147.
\newblock \url{https://ui.adsabs.harvard.edu/abs/1978MNRAS.182..147B/abstract}

\bibitem[{{Bell}(2004)}]{bell04}
---. 2004, MNRAS, 353, 550, \dodoi{10.1111/j.1365-2966.2004.08097.x}

\bibitem[{{Berezhko} \& {V{\"o}lk}(2007)}]{berezhko+07}
{Berezhko}, E.~G., \& {V{\"o}lk}, H.~J. 2007, \apjl, 661, L175, \dodoi{10.1086/518737}

\bibitem[{{Bisnovatyi-Kogan} \& {Silich}(1995)}]{bisnovatyi-kogan+95}
{Bisnovatyi-Kogan}, G.~S., \& {Silich}, S.~A. 1995, Reviews of Modern Physics, 67, 661, \dodoi{10.1103/RevModPhys.67.661}

\bibitem[{{Blandford} \& {Ostriker}(1978)}]{blandford+78}
{Blandford}, R.~D., \& {Ostriker}, J.~P. 1978, ApJL, 221, L29, \dodoi{10.1086/182658}

\bibitem[{{Blasi}(2002)}]{blasi02}
{Blasi}, P. 2002, APh, 16, 429

\bibitem[{{Blasi}(2004)}]{blasi04}
---. 2004, APh, 21, 45, \dodoi{10.1016/j.astropartphys.2003.10.008}

\bibitem[{Blondin {et~al.}(1998)Blondin, Wright, Borkowski, \& Reynolds}]{blondin+98}
Blondin, J.~M., Wright, E.~B., Borkowski, K.~J., \& Reynolds, S.~P. 1998, The Astrophysical Journal, 500, 342, \dodoi{10.1086/305708}

\bibitem[{{Brose, R.} {et~al.}(2020){Brose, R.}, {Pohl, M.}, {Sushch, I.}, {Petruk, O.}, \& {Kuzyo, T.}}]{brose+20}
{Brose, R.}, {Pohl, M.}, {Sushch, I.}, {Petruk, O.}, \& {Kuzyo, T.} 2020, A\&A, 634, A59, \dodoi{10.1051/0004-6361/201936567}

\bibitem[{{Bykov} {et~al.}(2013){Bykov}, {Brandenburg}, {Malkov}, \& {Osipov}}]{bykov+13}
{Bykov}, A.~M., {Brandenburg}, A., {Malkov}, M.~A., \& {Osipov}, S.~M. 2013, \ssr, 178, 201, \dodoi{10.1007/s11214-013-9988-3}

\bibitem[{{Caprioli}(2012)}]{caprioli12}
{Caprioli}, D. 2012, \jcap, 7, 38, \dodoi{10.1088/1475-7516/2012/07/038}

\bibitem[{{Caprioli} {et~al.}(2010{\natexlab{a}}){Caprioli}, {Amato}, \& {Blasi}}]{caprioli+10a}
{Caprioli}, D., {Amato}, E., \& {Blasi}, P. 2010{\natexlab{a}}, APh, 33, 160, \dodoi{10.1016/j.astropartphys.2010.01.002}

\bibitem[{{Caprioli} {et~al.}(2010{\natexlab{b}}){Caprioli}, {Amato}, \& {Blasi}}]{caprioli+10b}
---. 2010{\natexlab{b}}, APh, 33, 307, \dodoi{10.1016/j.astropartphys.2010.03.001}

\bibitem[{{Caprioli} {et~al.}(2009){Caprioli}, {Blasi}, {Amato}, \& {Vietri}}]{caprioli+09a}
{Caprioli}, D., {Blasi}, P., {Amato}, E., \& {Vietri}, M. 2009, MNRAS, 395, 895, \dodoi{10.1111/j.1365-2966.2009.14570.x}

\bibitem[{{Caprioli} {et~al.}(2015){Caprioli}, {Pop}, \& {Spitkovsky}}]{caprioli+15}
{Caprioli}, D., {Pop}, A., \& {Spitkovsky}, A. 2015, \apjl, 798, 28.
\newblock \doarXiv{1409.8291}

\bibitem[{{Caprioli} \& {Spitkovsky}(2014{\natexlab{a}})}]{caprioli+14a}
{Caprioli}, D., \& {Spitkovsky}, A. 2014{\natexlab{a}}, \apj, 783, 91, \dodoi{10.1088/0004-637X/783/2/91}

\bibitem[{{Caprioli} \& {Spitkovsky}(2014{\natexlab{b}})}]{caprioli+14c}
---. 2014{\natexlab{b}}, \apj, 794, 47, \dodoi{10.1088/0004-637X/794/1/47}

\bibitem[{Cardillo {et~al.}(2016)Cardillo, Amato, \& Blasi}]{cardillo+16}
Cardillo, M., Amato, E., \& Blasi, P. 2016, \aap, 595, A58, \dodoi{10.1051/0004-6361/201628669}

\bibitem[{{Castelletti} {et~al.}(2007){Castelletti}, {Dubner}, {Brogan}, \& {Kassim}}]{castelletti+07}
{Castelletti}, G., {Dubner}, G., {Brogan}, C., \& {Kassim}, N.~E. 2007, \aap, 471, 537, \dodoi{10.1051/0004-6361:20077062}

\bibitem[{{Castelletti} {et~al.}(2011){Castelletti}, {Dubner}, {Clarke}, \& {Kassim}}]{castelletti+11}
{Castelletti}, G., {Dubner}, G., {Clarke}, T., \& {Kassim}, N.~E. 2011, \aap, 534, A21, \dodoi{10.1051/0004-6361/201016081}

\bibitem[{{Chevalier}(1982)}]{chevalier82}
{Chevalier}, R.~A. 1982, \apj, 258, 790, \dodoi{10.1086/160126}

\bibitem[{{Chevalier} \& {Gardner}(1974)}]{chevalier+74}
{Chevalier}, R.~A., \& {Gardner}, J. 1974, \apj, 192, 457, \dodoi{10.1086/153077}

\bibitem[{Cioffi {et~al.}(1988)Cioffi, McKee, \& Bertschinger}]{cioffi+88}
Cioffi, D.~F., McKee, C.~F., \& Bertschinger, E. 1988, \apj, 334, 252, \dodoi{10.1086/166834}

\bibitem[{{Cotton} {et~al.}(2024){Cotton}, {Kothes}, {Camilo}, {Chandra}, {Buchner}, \& {Nyamai}}]{cotton+24}
{Cotton}, W.~D., {Kothes}, R., {Camilo}, F., {et~al.} 2024, \apjs, 270, 21, \dodoi{10.3847/1538-4365/ad0ecb}

\bibitem[{Crain \& van~de Voort(2023)}]{crain+23}
Crain, R.~A., \& van~de Voort, F. 2023, Annual Review of Astronomy and Astrophysics, 61, 473, \dodoi{https://doi.org/10.1146/annurev-astro-041923-043618}

\bibitem[{{Cristofari} {et~al.}(2021){Cristofari}, {Blasi}, \& {Caprioli}}]{cristofari+21}
{Cristofari}, P., {Blasi}, P., \& {Caprioli}, D. 2021, \aap, 650, A62, \dodoi{10.1051/0004-6361/202140448}

\bibitem[{{Diesing}(2023)}]{diesing23}
{Diesing}, R. 2023, arXiv e-prints, arXiv:2305.07697, \dodoi{10.48550/arXiv.2305.07697}

\bibitem[{{Diesing} \& {Caprioli}(2018)}]{diesing+18}
{Diesing}, R., \& {Caprioli}, D. 2018, Physical Review Letters, 121, 091101, \dodoi{10.1103/PhysRevLett.121.091101}

\bibitem[{{Diesing} \& {Caprioli}(2019)}]{diesing+19}
---. 2019, \prl, 123, 071101, \dodoi{10.1103/PhysRevLett.123.071101}

\bibitem[{{Diesing} \& {Caprioli}(2021)}]{diesing+21}
---. 2021, \apj, 922, 1, \dodoi{10.3847/1538-4357/ac22fe}

\bibitem[{Diesing {et~al.}(2024)Diesing, Guo, Kim, Stone, \& Caprioli}]{diesing+24}
Diesing, R., Guo, M., Kim, C.-G., Stone, J., \& Caprioli, D. 2024, The Astrophysical Journal, 974, 201, \dodoi{10.3847/1538-4357/ad74f0}

\bibitem[{Draine(2011)}]{draine11}
Draine, B.~T. 2011, Physics of the Interstellar and Intergalactic Medium.
\newblock \url{http://adsabs.harvard.edu/abs/2011piim.book.....D}

\bibitem[{{El-Badry} {et~al.}(2019){El-Badry}, {Ostriker}, {Kim}, {Quataert}, \& {Weisz}}]{el-badry+19}
{El-Badry}, K., {Ostriker}, E.~C., {Kim}, C.-G., {Quataert}, E., \& {Weisz}, D.~R. 2019, \mnras, 490, 1961, \dodoi{10.1093/mnras/stz2773}

\bibitem[{{Feldmann} {et~al.}(2023){Feldmann}, {Quataert}, {Faucher-Gigu{\`e}re}, {Hopkins}, {{\c{C}}atmabacak}, {Kere{\v{s}}}, {Bassini}, {Bernardini}, {Bullock}, {Cenci}, {Gensior}, {Liang}, {Moreno}, \& {Wetzel}}]{feldmann+23}
{Feldmann}, R., {Quataert}, E., {Faucher-Gigu{\`e}re}, C.-A., {et~al.} 2023, \mnras, 522, 3831, \dodoi{10.1093/mnras/stad1205}

\bibitem[{{Fermi}(1954)}]{fermi54}
{Fermi}, E. 1954, Ap. J., 119, 1, \dodoi{10.1086/145789}

\bibitem[{{Guo} {et~al.}(2024){Guo}, {Kim}, \& {Stone}}]{guo+24}
{Guo}, M., {Kim}, C.-G., \& {Stone}, J.~M. 2024, arXiv e-prints, arXiv:2411.12809.
\newblock \doarXiv{2411.12809}

\bibitem[{{Gupta} {et~al.}(2024){Gupta}, {Caprioli}, \& {Spitkovsky}}]{gupta+24b}
{Gupta}, S., {Caprioli}, D., \& {Spitkovsky}, A. 2024, arXiv e-prints, arXiv:2408.16071, \dodoi{10.48550/arXiv.2408.16071}

\bibitem[{{Gupta} {et~al.}(2018){Gupta}, {Nath}, \& {Sharma}}]{gupta+18}
{Gupta}, S., {Nath}, B.~B., \& {Sharma}, P. 2018, \mnras, 479, 5220, \dodoi{10.1093/mnras/sty1846}

\bibitem[{{Gupta} {et~al.}(2020){Gupta}, {Nath}, {Sharma}, \& {Eichler}}]{gupta+20}
{Gupta}, S., {Nath}, B.~B., {Sharma}, P., \& {Eichler}, D. 2020, \mnras, 493, 3159, \dodoi{10.1093/mnras/staa286}

\bibitem[{{Gupta} {et~al.}(2016){Gupta}, {Nath}, {Sharma}, \& {Shchekinov}}]{gupta+16}
{Gupta}, S., {Nath}, B.~B., {Sharma}, P., \& {Shchekinov}, Y. 2016, \mnras, 462, 4532, \dodoi{10.1093/mnras/stw1920}

\bibitem[{{Gupta} {et~al.}(2021){Gupta}, {Sharma}, \& {Mignone}}]{gupta+21a}
{Gupta}, S., {Sharma}, P., \& {Mignone}, A. 2021, \mnras, 502, 2733, \dodoi{10.1093/mnras/stab142}

\bibitem[{{H.~E.~S.~S. Collaboration} {et~al.}(2018){H.~E.~S.~S. Collaboration}, {Abdalla}, {Abramowski}, {Aharonian}, {Ait Benkhali}, {Ang{\"u}ner}, {Arakawa}, {Arrieta}, {Aubert}, {Backes}, {Balzer}, {Barnard}, {Becherini}, {Becker Tjus}, {Berge}, {Bernhard}, {Bernl{\"o}hr}, {Blackwell}, {B{\"o}ttcher}, {Boisson}, {Bolmont}, {Bonnefoy}, {Bordas}, {Bregeon}, {Brun}, {Brun}, {Bryan}, {B{\"u}chele}, {Bulik}, {Capasso}, {Caroff}, {Carosi}, {Casanova}, {Cerruti}, {Chakraborty}, {Chaves}, {Chen}, {Chevalier}, {Colafrancesco}, {Condon}, {Conrad}, {Davids}, {Decock}, {Deil}, {Devin}, {deWilt}, {Dirson}, {Djannati-Ata{\"\i}}, {Donath}, {Drury}, {Dutson}, {Dyks}, {Edwards}, {Egberts}, {Emery}, {Ernenwein}, {Eschbach}, {Farnier}, {Fegan}, {Fernandes}, {Fernandez}, {Fiasson}, {Fontaine}, {Funk}, {F{\"u}{\ss}ling}, {Gabici}, {Gallant}, {Garrigoux}, {Gat{\'e}}, {Giavitto}, {Giebels}, {Glawion}, {Glicenstein}, {Gottschall}, {Grondin}, {Hahn}, {Haupt}, {Hawkes}, {Heinzelmann}, {Henri}, {Hermann}, {Hinton}, {Hofmann},
  {Hoischen}, {Holch}, {Holler}, {Horns}, {Ivascenko}, {Iwasaki}, {Jacholkowska}, {Jamrozy}, {Jankowsky}, {Jankowsky}, {Jingo}, {Jouvin}, {Jung-Richardt}, {Kastendieck}, {Katarzy{\'n}ski}, {Katsuragawa}, {Katz}, {Kerszberg}, {Khangulyan}, {Kh{\'e}lifi}, {King}, {Klepser}, {Klochkov}, {Klu{\'z}niak}, {Komin}, {Kosack}, {Krakau}, {Kraus}, {Kr{\"u}ger}, {Laffon}, {Lamanna}, {Lau}, {Lees}, {Lefaucheur}, {Lemi{\`e}re}, {Lemoine-Goumard}, {Lenain}, {Leser}, {Lohse}, {Lorentz}, {Liu}, {L{\'o}pez-Coto}, {Lypova}, {Malyshev}, {Marandon}, {Marcowith}, {Mariaud}, {Marx}, {Maurin}, {Maxted}, {Mayer}, {Meintjes}, {Meyer}, {Mitchell}, {Moderski}, {Mohamed}, {Mohrmann}, {Mor{\r{a}}}, {Moulin}, {Murach}, {Nakashima}, {de Naurois}, {Ndiyavala}, {Niederwanger}, {Niemiec}, {Oakes}, {O'Brien}, {Odaka}, {Ohm}, {Ostrowski}, {Oya}, {Padovani}, {Panter}, {Parsons}, {Pekeur}, {Pelletier}, {Perennes}, {Petrucci}, {Peyaud}, {Piel}, {Pita}, {Poireau}, {Poon}, {Prokhorov}, {Prokoph}, {P{\"u}hlhofer}, {Punch}, {Quirrenbach}, {Raab},
  {Rauth}, {Reimer}, {Reimer}, {Renaud}, {de los Reyes}, {Rieger}, {Rinchiuso}, {Romoli}, {Rowell}, {Rudak}, {Rulten}, {Safi-Harb}, {Sahakian}, {Saito}, {Sanchez}, {Santangelo}, {Sasaki}, {Schlickeiser}, {Sch{\"u}ssler}, {Schulz}, {Schwanke}, {Schwemmer}, {Seglar-Arroyo}, {Settimo}, {Seyffert}, {Shafi}, {Shilon}, {Shiningayamwe}, {Simoni}, {Sol}, {Spanier}, {Spir-Jacob}, {Stawarz}, {Steenkamp}, {Stegmann}, {Steppa}, {Sushch}, {Takahashi}, {Tavernet}, {Tavernier}, {Taylor}, {Terrier}, {Tibaldo}, {Tiziani}, {Tluczykont}, {Trichard}, {Tsirou}, {Tsuji}, {Tuffs}, {Uchiyama}, {van der Walt}, {van Eldik}, {van Rensburg}, {van Soelen}, {Vasileiadis}, {Veh}, {Venter}, {Viana}, {Vincent}, {Vink}, {Voisin}, {V{\"o}lk}, {Vuillaume}, {Wadiasingh}, {Wagner}, {Wagner}, {Wagner}, {White}, {Wierzcholska}, {Willmann}, {W{\"o}rnlein}, {Wouters}, {Yang}, {Zaborov}, {Zacharias}, {Zanin}, {Zdziarski}, {Zech}, {Zefi}, {Ziegler}, {Zorn}, \& {{\.Z}ywucka}}]{HESS18b}
{H.~E.~S.~S. Collaboration}, {Abdalla}, H., {Abramowski}, A., {et~al.} 2018, \aap, 612, A3, \dodoi{10.1051/0004-6361/201732125}

\bibitem[{{Haggerty} \& {Caprioli}(2020)}]{haggerty+20}
{Haggerty}, C.~C., \& {Caprioli}, D. 2020, \apj, 905, 1, \dodoi{10.3847/1538-4357/abbe06}

\bibitem[{Heckman \& Thompson(2017)}]{ht17}
Heckman, T.~M., \& Thompson, T.~A. 2017, ArXiv e-prints.
\newblock \doarXiv{1701.09062}

\bibitem[{{Hillas}(2005)}]{hillas05}
{Hillas}, A.~M. 2005, Journal of Physics G Nuclear Physics, 31, 95, \dodoi{10.1088/0954-3899/31/5/R02}

\bibitem[{Hopkins {et~al.}(2018)Hopkins, Wetzel, Kere{\v s}, Faucher-Gigu{\`e}re, Quataert, Boylan-Kolchin, Murray, Hayward, \& El-Badry}]{hopkins+18}
Hopkins, P.~F., Wetzel, A., Kere{\v s}, D., {et~al.} 2018, \mnras, 477, 1578, \dodoi{10.1093/mnras/sty674}

\bibitem[{Kim \& Ostriker(2015)}]{kim+15}
Kim, C.-G., \& Ostriker, E.~C. 2015, \apj, 802, 99, \dodoi{10.1088/0004-637X/802/2/99}

\bibitem[{Kobashi {et~al.}(2022)Kobashi, Yasuda, \& Lee}]{kobashi+22}
Kobashi, R., Yasuda, H., \& Lee, S.-H. 2022, The Astrophysical Journal, 936, 26, \dodoi{10.3847/1538-4357/ac80f9}

\bibitem[{{Koo} {et~al.}(2020){Koo}, {Kim}, {Park}, \& {Ostriker}}]{koo+20}
{Koo}, B.-C., {Kim}, C.-G., {Park}, S., \& {Ostriker}, E.~C. 2020, \apj, 905, 35, \dodoi{10.3847/1538-4357/abc1e7}

\bibitem[{{Krymskii}(1977)}]{krymskii77}
{Krymskii}, G.~F. 1977, Akademiia Nauk SSSR Doklady, 234, 1306.
\newblock \url{https://ui.adsabs.harvard.edu/abs/1977DoSSR.234R1306K}

\bibitem[{Kulsrud \& Pearce(1968)}]{kulsrud+68}
Kulsrud, R., \& Pearce, W. 1968, The Astronomical Journal Supplement, 73, 22

\bibitem[{{Lagage} \& {Cesarsky}({1983a})}]{lagage+83a}
{Lagage}, P.~O., \& {Cesarsky}, C.~J. {1983a}, A\&A, 118, 223.
\newblock \url{http://adsabs.harvard.edu/abs/1983A26A...118..223L}

\bibitem[{{Lee} {et~al.}(2015){Lee}, {Patnaude}, {Raymond}, {Nagataki}, {Slane}, \& {Ellison}}]{lee+15}
{Lee}, S.-H., {Patnaude}, D.~J., {Raymond}, J.~C., {et~al.} 2015, \apj, 806, 71, \dodoi{10.1088/0004-637X/806/1/71}

\bibitem[{{Malkov}(1997)}]{malkov97}
{Malkov}, M.~A. 1997, Ap. J., 485, 638, \dodoi{10.1086/304471}

\bibitem[{{Malkov} {et~al.}(2000){Malkov}, {Diamond}, \& {V{\"o}lk}}]{malkov+00}
{Malkov}, M.~A., {Diamond}, P.~H., \& {V{\"o}lk}, H.~J. 2000, Ap. J. L., 533, L171, \dodoi{10.1086/312622}

\bibitem[{Mignone {et~al.}(2007)Mignone, Bodo, Massaglia, Matsakos, Tesileanu, Zanni, \& Ferrari}]{mignone+07}
Mignone, A., Bodo, G., Massaglia, S., {et~al.} 2007, \apjs, 170, 228, \dodoi{10.1086/513316}

\bibitem[{{Morlino} \& {Caprioli}(2012)}]{morlino+12}
{Morlino}, G., \& {Caprioli}, D. 2012, A\&A, 538, A81, \dodoi{10.1051/0004-6361/201117855}

\bibitem[{{Ostriker} \& {McKee}(1988)}]{ostriker+88}
{Ostriker}, J.~P., \& {McKee}, C.~F. 1988, Reviews of Modern Physics, 60, 1, \dodoi{10.1103/RevModPhys.60.1}

\bibitem[{{Park} {et~al.}(2015){Park}, {Caprioli}, \& {Spitkovsky}}]{park+15}
{Park}, J., {Caprioli}, D., \& {Spitkovsky}, A. 2015, Physical Review Letters, 114, 085003, \dodoi{10.1103/PhysRevLett.114.085003}

\bibitem[{{Petruk} {et~al.}(2018){Petruk}, {Kuzyo}, {Orlando}, {Pohl}, {Miceli}, {Bocchino}, {Beshley}, \& {Brose}}]{petruk+18}
{Petruk}, O., {Kuzyo}, T., {Orlando}, S., {et~al.} 2018, \mnras, 479, 4253, \dodoi{10.1093/mnras/sty1750}

\bibitem[{Pfrommer {et~al.}(2017)Pfrommer, Pakmor, Schaal, Simpson, \& Springel}]{pfrommer+17}
Pfrommer, C., Pakmor, R., Schaal, K., Simpson, C.~M., \& Springel, V. 2017, \mnras, 465, 4500, \dodoi{10.1093/mnras/stw2941}

\bibitem[{Pillepich {et~al.}(2018)Pillepich, Springel, Nelson, Genel, Naiman, Pakmor, Hernquist, Torrey, Vogelsberger, Weinberger, \& Marinacci}]{pillepich+18}
Pillepich, A., Springel, V., Nelson, D., {et~al.} 2018, \mnras, 473, 4077, \dodoi{10.1093/mnras/stx2656}

\bibitem[{{Ptuskin} {et~al.}(2010){Ptuskin}, {Zirakashvili}, \& {Seo}}]{ptuskin+10}
{Ptuskin}, V., {Zirakashvili}, V., \& {Seo}, E.-S. 2010, \apj, 718, 31, \dodoi{10.1088/0004-637X/718/1/31}

\bibitem[{{Rodr{\'\i}guez Montero} {et~al.}(2022){Rodr{\'\i}guez Montero}, {Martin-Alvarez}, {Sijacki}, {Slyz}, {Devriendt}, \& {Dubois}}]{montero+22}
{Rodr{\'\i}guez Montero}, F., {Martin-Alvarez}, S., {Sijacki}, D., {et~al.} 2022, \mnras, 511, 1247, \dodoi{10.1093/mnras/stab3716}

\bibitem[{{Sedov}(1959)}]{sedov59}
{Sedov}, L.~I. 1959, {Similarity and Dimensional Methods in Mechanics}

\bibitem[{{Sharma} {et~al.}(2014){Sharma}, {Roy}, {Nath}, \& {Shchekinov}}]{sharma+14}
{Sharma}, P., {Roy}, A., {Nath}, B.~B., \& {Shchekinov}, Y. 2014, \mnras, 443, 3463, \dodoi{10.1093/mnras/stu1307}

\bibitem[{{Skilling}(1975)}]{skilling75a}
{Skilling}, J. 1975, MNRAS, 172, 557.
\newblock \url{http://adsabs.harvard.edu/abs/1975MNRAS.172..557S}

\bibitem[{{Skilling}({1975b})}]{skilling75b}
---. {1975b}, MNRAS, 173, 245.
\newblock \url{http://adsabs.harvard.edu/abs/1975MNRAS.173..245S}

\bibitem[{{Skilling}({1975c})}]{skilling75c}
---. {1975c}, MNRAS, 173, 255.
\newblock \url{http://adsabs.harvard.edu/abs/1975MNRAS.173..255S}

\bibitem[{Slane {et~al.}(2014)Slane, Lee, Ellison, Patnaude, Hughes, Eriksen, Castro, \& Nagataki}]{slane+14}
Slane, P., Lee, S.-H., Ellison, D.~C., {et~al.} 2014, \apj, 783, 33, \dodoi{10.1088/0004-637X/783/1/33}

\bibitem[{{Taylor}(1950)}]{taylor50}
{Taylor}, G. 1950, Proceedings of the Royal Society of London Series A, 201, 159, \dodoi{10.1098/rspa.1950.0049}

\bibitem[{{Truelove} \& {Mc Kee}(1999)}]{truelove+99}
{Truelove}, J.~K., \& {Mc Kee}, C.~F. 1999, ApJ~Supplement Series, 120, 299, \dodoi{10.1086/313176}

\bibitem[{{Vishniac}(1983)}]{vishniac83}
{Vishniac}, E.~T. 1983, \apj, 274, 152, \dodoi{10.1086/161433}

\bibitem[{{Yadav} {et~al.}(2017){Yadav}, {Mukherjee}, {Sharma}, \& {Nath}}]{yadav+17}
{Yadav}, N., {Mukherjee}, D., {Sharma}, P., \& {Nath}, B.~B. 2017, \mnras, 465, 1720, \dodoi{10.1093/mnras/stw2522}

\bibitem[{{Zabalza}(2015)}]{naima}
{Zabalza}, V. 2015, Proc.~of International Cosmic Ray Conference 2015, 922

\bibitem[{{Zacharegkas} {et~al.}(2022){Zacharegkas}, {Caprioli}, \& {Haggerty}}]{zacharegkas+22}
{Zacharegkas}, G., {Caprioli}, D., \& {Haggerty}, C. 2022, arXiv e-prints, arXiv:2210.08072.
\newblock \doarXiv{2210.08072}

\bibitem[{{Zacharegkas} {et~al.}(2024){Zacharegkas}, {Caprioli}, {Haggerty}, {Gupta}, \& {Schroer}}]{zacharegkas+24}
{Zacharegkas}, G., {Caprioli}, D., {Haggerty}, C., {Gupta}, S., \& {Schroer}, B. 2024, \apj, 967, 71, \dodoi{10.3847/1538-4357/ad3960}

\bibitem[{Zirakashvili \& Aharonian(2007)}]{zirakashvili+07}
Zirakashvili, V.~N., \& Aharonian, F. 2007, A\&A, 465, 695, \dodoi{10.1051/0004-6361:20066494}

\bibitem[{{Zweibel}(1979)}]{zweibel79}
{Zweibel}, E.~G. 1979, in American Institute of Physics Conference Series, Vol.~56, Particle Acceleration Mechanisms in Astrophysics, ed. J.~{Arons}, C.~{McKee}, \& C.~{Max}, 319--328, \dodoi{10.1063/1.32090}

\end{thebibliography}

\end{document}